\documentclass[sigconf]{acmart}
\usepackage{acmart-taps}
\usepackage{subcaption}
\usepackage{amsfonts}
\usepackage{tabularx}
\usepackage{microtype}
\usepackage{stfloats}
\usepackage{balance}
\usepackage{xcolor}
\usepackage{float}
\usepackage{multirow}
\usepackage{graphicx}
\usepackage{enumitem}

\usepackage[T1]{fontenc}
\usepackage{xparse}
\usepackage{enumitem}

\usepackage{chi25-1034}

\aptLtoXcmd{
\def\User{\textbf{User:\xspace}}
\def\Chatbot{\textbf{Chatbot:\xspace}}
\def\Removed{\textbf{Removed:\xspace}}}{}

\newcommand{\iQuote}[1]{{``\textit{#1}''}}
\newcommand{\eg}{\textit{e.g.}, }
\newcommand{\ie}{\textit{i.e.}, }

\newcommand{\tdescriptive}{\textsc{Descriptive}}
\newcommand{\tidiographic}{\textsc{Idiographic}}
\newcommand{\tclarifying}{\textsc{Clarifying}}
\newcommand{\texplanatory}{\textsc{Explanatory}}

\newcommand{\texploration}{\textsc{Exploration}}
\newcommand{\trequirements}{\textsc{Requirements}}
\newcommand{\tevaluation}{\textsc{Evaluation}}

\newcolumntype{A}{>{\hsize=.07\hsize}X}
\newcolumntype{B}{>{\hsize=.30\hsize}X}
\newcolumntype{C}{>{\hsize=.12\hsize}X}
\newcolumntype{D}{>{\hsize=.44\hsize}X}

\newcolumntype{E}{>{\hsize=.12\hsize}X}
\newcolumntype{F}{>{\hsize=.23\hsize}X}
\newcolumntype{G}{>{\hsize=.65\hsize}X}

\newcolumntype{S}{>{\hsize=.08\hsize}X}

\newcolumntype{H}{>{\hsize=.07\hsize}X}
\newcolumntype{I}{>{\hsize=.82\hsize}X}
\newcolumntype{J}{>{\hsize=.04\hsize}X}
\newcolumntype{K}{>{\hsize=.07\hsize}X}

\newcolumntype{L}{>{\hsize=.13\hsize}X}
\newcolumntype{M}{>{\hsize=.22\hsize}X}

\newcommand{\change}[1]{\textcolor{black}{#1}}

\AtBeginDocument{%
  \providecommand\BibTeX{{%
    \normalfont B\kern-0.5em{\scshape i\kern-0.25em b}\kern-0.8em\TeX}}}

\copyrightyear{2025}
\acmYear{2025}
\setcopyright{cc}
\setcctype{by}
\acmConference[CHI '25]{CHI Conference on Human Factors in Computing Systems}{April 26-May 1, 2025}{Yokohama, Japan}
\acmBooktitle{CHI Conference on Human Factors in Computing Systems (CHI '25), April 26-May 1, 2025, Yokohama, Japan}\acmDOI{10.1145/3706598.3714128}
\acmISBN{979-8-4007-1394-1/25/04}

\begin{document}


\title[Chatbots for Data Collection in Surveys]{Chatbots for Data Collection in Surveys: \\A~Comparison~of~Four~Theory-Based~Interview~Probes}


\author{Rune M. Jacobsen}
\email{runemj@cs.aau.dk}
\orcid{0000-0002-1877-1845}
\affiliation{%
  \institution{Aalborg University}
  \city{Aalborg}
  \country{Denmark}
}

\author{Samuel Rhys Cox}
\email{srcox@cs.aau.dk}
\orcid{0000-0002-4558-6610}
\affiliation{%
  \institution{Aalborg University}
  \city{Aalborg}
  \country{Denmark}
}

\author{Carla F. Griggio}
\email{cfg@cs.aau.dk}
\orcid{0000-0001-9133-3828}
\affiliation{%
  \institution{Aalborg University}
  \city{Copenhagen}
  \country{Denmark}
}

\author{Niels van Berkel}
\email{nielsvanberkel@cs.aau.dk}
\orcid{0000-0001-5106-7692}
\affiliation{%
  \institution{Aalborg University}
  \city{Aalborg}
  \country{Denmark}
}

\renewcommand{\shortauthors}{Jacobsen et al.}

\begin{abstract}

Surveys are a widespread method for collecting data at scale, but their rigid structure often limits the depth of qualitative insights obtained. While interviews naturally yield richer responses, they are challenging to conduct across diverse locations and large participant pools. To partially bridge this gap, we investigate the potential of using LLM-based chatbots to support qualitative data collection through interview probes embedded in surveys. We assess four theory-based interview probes: descriptive, idiographic, clarifying, and explanatory. Through a split-plot study design (\textit{N}~=~64), we compare the probes' impact on response quality and user experience across three key stages of HCI research: \textit{exploration}, \textit{requirements gathering}, and \textit{evaluation}. Our results show that probes facilitate the collection of high-quality survey data, with specific probes proving effective at different research stages. We contribute practical and methodological implications for using chatbots as research tools to enrich qualitative data collection.


\end{abstract}





\maketitle

\balance


\section{Introduction}
Online surveys are widely used in Human-Computer Interaction (HCI) research as a data collection method. They can reach a wide population~\cite{Evans2018}, are cost-effective~\cite{Olson2020}, and responses are quickly collected~\cite{Couper2008}. HCI research has extensively relied on online surveys to generate insights about users' attitudes, perceptions, habits, awareness, and experiences with technology~\cite{Muller2015}. 
Online surveys can provide both quantitative and qualitative self-reported data. For example, surveys often include a combination of closed-ended questions with restricted response options, facilitating quantification and comparison~\cite{lazar_research_2017}, and open-ended questions for accessing richer, more detailed insights through respondents' own words~\cite{Reja2003, chen2017, Singer2017}.  

However, the quality and richness of the qualitative data collected through online surveys is often more limited than when interacting with participants directly, such as when conducting face-to-face interviews. 
In surveys, any follow-up questions must be predetermined, potentially missing out on clarifying a vague answer or expanding on an interesting insight. 
Additionally, participant engagement may be less reliable~\cite{Wright2017,heerwegh2008, roster2004, fricker2005}, resulting in low-quality~\cite{Reja2003} or nonsensical answers~\cite{Gadiraju2015}, or in omitting responses altogether. 
Beyond data quality, asking open-ended questions in surveys may cause survey fatigue~\cite{porter2004overcoming}, often ascribed to the cognitive load and the additional time and effort required, leading to \textit{survey taking fatigue} and participant dropout~\cite{Porter2004b}. 
These limitations motivate the exploration of novel interactive methods for data collection that are capable of preserving the time- and cost-effectiveness provided by online surveys without compromising on the richness and quality of open-ended answers.

Recent research has explored the feasibility of collecting qualitative data through chatbots embedded in surveys, substituting open-ended questions with chatbot-based interview probes that interact with participants in a conversational style~\cite{Uka2021, Xiao2020, Ge2023, Kocielnik2021, Harms2017, Kim2019, Wijenayake2020ResearchBots}. 
Chatbot-driven data collection in surveys has been shown to enhance participant disclosure~\cite{Tallyn2018, Xiao2019}, improve data quality as compared to the traditional open-ended question format~\cite{Xiao2020, Kim2019}, and been hypothesised to reduce satisficing behaviour~\cite{Wijenayake2020ResearchBots}. 
However, chatbots often show limitations in understanding the diverse user input in responses to open-ended questions~\cite{devault2014}, leading to user disappointment and frustration~\cite{Grudin2019}. 
With the recent surge of large-language models (LLMs), chatbot capabilities have taken a leap in their ability to interpret text and provide a sensible response, proposing a promising direction for utilising chatbots in surveys. 

\change{We propose that LLM-based chatbots, instructed in conversational techniques based on face-to-face interviews, can facilitate the collection of high-quality qualitative data in surveys. A well-established approach to uncovering in-depth insights involves the use of interview probes, which aim to elicit richer and more nuanced insights based on participants' prior responses~\cite{Price2002, gorden1987interviewing, russel2000social, Robinson2023}. Incorporating such probing techniques in chatbot-driven survey inquiries presents an opportunity to enhance data quality. However, interview probes can also negatively affect the interviewee’s experience~\cite{willis2004cognitive, Priede03092014}. This underscores the importance of carefully considering the survey-takers’ experience when designing chatbots that employ probing questions.}
In this work, we are interested in exploring chatbot designs that can serve as tools for large-scale surveys to inform three common interview stages of HCI research and practice~\cite[pp. 180-187]{lazar_research_2017}: \texploration, \trequirements, and \tevaluation.  
To address these research gaps, we set out to answer two research questions:

\begin{description}[style=nextline, labelwidth=2em, labelsep=1em, leftmargin=3em]
    \item[RQ1] \textit{How can distinct chatbot-based interview probes support qualitative data collection across different interview stages of HCI research?}
    \item[RQ2] \textit{What is the user experience of interacting with a chatbot-based interview probe when completing a survey?}
    
\end{description}

To address these research questions, we incorporate and assess four theory-based interviewing probes~\cite{Robinson2023} in a chatbot for collecting high-quality qualitative data through online surveys. We examine the effectiveness of a \tdescriptive, \tidiographic, \tclarifying, and \texplanatory~interview probes in a between-subject setup across the common HCI research interview stages of \texploration, \trequirements, and \tevaluation~\cite[pp. 180-187]{lazar_research_2017} in a within-subjects design. 
Our analysis focuses on assessing the impact of each interviewing probe on the quality of responses, using Gricean Maxims (a set of communication principles) to measure the informativeness, relevance, specificity, and clarity, as well as the participants' experience. 

We contribute an analysis of the effect of interview probes and research interview stages on response quality, as well as participant experience and perception of survey chatbots.
Furthermore, we provide concrete recommendations for employing interview probes in HCI studies and outline opportunities and challenges for integrating LLM-based chatbots in surveys. Finally, we provide the source code and implementation details of our chatbot-based interview probes to support other researchers seeking to employ chatbots for data collection in surveys.




\section{Background and Related Work}
\subsection{Information Elicitation through Online Surveys}
Conducting surveys is one of the primary methods for collecting insights into the opinions, experiences and perceptions of people~\cite{Evans2005}. In surveys, researchers and practitioners utilise a series of questions to elicit information on a topic of interest which is either closed- or open-ended. Open-ended questions are crucial in collecting rich and nuanced data from participants~\cite{Reja2003}, and in addition, attempt to uncover the \textit{why} of participant responses to closed-ended questions~\cite{chen2017, Singer2017}. Several factors can reduce the quality of responses to open-ended questions. This includes survey fatigue~\cite{porter2004overcoming}, where participants feel overwhelmed by the questions posed~\cite{Lavrakas2008, Porter2004a} and either drop out or provide lower-quality responses by responding faster~\cite{Xiao2020, Porter2004b}. Furthermore, answering open-ended questions is regarded as taking extra time and demanding increased effort due to the participant formulating their own words and typing out their responses~\cite{chen2017}. These factors can lead to participants skipping this type of question or even providing gibberish responses~\cite{Gadiraju2015}. In addition, researchers argue that it can be challenging to manage the progress of web surveys compared to in-person surveys~\cite{heerwegh2008, roster2004, fricker2005}, but that measurement errors are less prevalent when compared to surveys conducted in-person or over the phone.

Researchers have introduced various techniques to counteract the drawbacks (such as reduced response quality) of open-ended questions. Web surveys have previously integrated lightweight probes, subtly modifying open-ended questions based on, for example, prior scale responses~\cite{oudejans2018using, Behr2012}. In addition, researchers have created personalised surveys that incorporate prior participant data (\eg social media and health trackers) into the formulation of survey questions~\cite{Velykoivanenko2024}. In our work, we specifically focus on the systematic posing of follow-up questions based on immediate prior participant responses.

\subsection{Conversational AI and Chatbots}
Introducing conversational elements as a means to elicit information has been explored in a variety of contexts, since conversational interfaces possess several advantages over traditional WIMP (Windows, Icons, Menus and Pointers) interfaces~\cite{Luger2016, traum2017computational}, for instance, they do not lock users into a fixed path when users provide diverse requests~\cite{traum2017computational}. This is why researchers and practitioners have deployed chatbots in myriad scenarios. For instance, customer service agents~\cite{celino2020submitting}, tutoring students~\cite{graesser2005}, planning and scheduling~\cite{cranshaw2017, Hemphill1990, McGlashan1992}, recommending products, jobs or movies~\cite{Kang2017, Zhang2018}, personal assistants~\cite{Liao2018}, psycho therapy~\cite{Schroeder2018}, group decision support~\cite{Shamekhi2018}, support in voting~\cite{folstad2017}, workers companion~\cite{Williams2018}. Chatbots can enable gathering information from users both in general~\cite{Radlinski2017, Ricci2022, Trippas2018} and for specific tasks such as making recommendations~\cite{Xiao2020}, and for doing specific tasks~\cite{bohus2009}.

In the context of surveys, prior work posits that chatbots can potentially improve the qualitative data collection process. This includes explaining the purpose of a question when asked by a participant and guiding them~\cite{Li2017}. A chatbot has the ability to present survey questions as personalised, conversational messages, potentially enhancing participant involvement and the quality of their responses~\cite{Heerwegh2007, Krosnick1999}. Kim et al. found that, through satisficing theory, chatbots elicit higher quality responses compared to traditional surveys~\cite{Kim2019}, which is supported by~\citet{Xiao2020}, who furthermore found that a chatbot may increase participant engagement. Previous studies have shown that the social behaviours displayed by agents can successfully enhance user engagement across different social settings, as measured by indicators such as the duration of the interaction, the extent and depth of personal information shared, and a positive perception of the agent and the results of the interaction~\cite{Bickmore2011, Shamekhi2018}. In addition, participants have been found willing to share sensitive and personal information with a chatbot~\cite{Sundar2019}. 

A key factor in utilising chatbots as part of web surveys is to evaluate their efficacy in eliciting information. The evaluation of conversational interfaces has traditionally been divided into two different directions, objective~\cite{Dybkjaer2004, liu2017, Walker1997} and subjective metrics~\cite{Hone2000, Bartneck2009, Bickmore2001, Novick2014}. Objective metrics are calculated directly from interaction logs and may include measures such as task or domain coverage, error rate, number of issues encountered during the interaction, accuracy, or similar metrics which can be benchmarked~\cite{Dybkjaer2004, liu2017, Walker1997}. It may also include calculating task achievement, dialogue efficiency (such as duration and total number of exchanges), and dialogue quality (\eg response time)~\cite{Walker1997}. Subjective metrics typically rely on user feedback regarding specific elements, such as satisfaction and comprehensibility (\eg ~\cite{Hone2000}). For HCI in particular, researchers have evaluated chatbot characteristics and behaviour such as anthropomorphism and likeability~\cite{Bartneck2009}, ability to build rapport~\cite{Bickmore2001, Novick2014}, and trust~\cite{Bickmore2001}. 

In this work, we set out to utilise chatbots as interactive data collection in surveys, given their proven ability to improve data quality, participant engagement and disclosure. 






\subsection{Interviewing Probes} \label{sec:probes}
Research interviews in social sciences have evolved notably over the past four decades, highlighted by significant contributions from Gorden~\cite{gorden1987interviewing}, Mishler~\cite{mishler1991research}, Wengraf~\cite{Wengraf2001}, Roulston~\cite{Roulston2014}, and Brinkmann and Kvale~\cite{brinkman2018}. Probing, essential for uncovering rich, nuanced data, plays a key role in this evolution. Probing describes the asking of strategic follow-up questions that reveal deeper insights into participants' experiences, progressively uncovering potential hidden layers of participants' experiences or insights~\cite{Price2002}. 
Gorden~\cite{gorden1987interviewing} and Bernard~\cite{russel2000social} have presented probing taxonomies. Gorden's six-probe framework aims to support interviewers eliciting comprehensive responses~\cite{gorden1987interviewing}, ranging from silent and encouragement probes to more specific ones like elaboration, clarification, recapitulation, and reflective probes. Bernard built on this framework, including directive and baiting probes for eliciting more disclosure~\cite{russel2000social}. 

\begin{table*}[b]
\caption{Hypothetical examples of the four interviewing probes within HCI.}
\label{tab:hciexamples}
\begin{tabularx}{\linewidth}{@{}l|XXXX@{}}
\toprule
{\textbf{Role}} & \textbf{Descriptive} & \textbf{Idiographic} & \textbf{Clarifying} & \textbf{Explanatory} \\ 
    \midrule
    Interviewer & \multicolumn{4}{c}{\textit{What is your experience with technology in your life?}} \\ \\
    Participant & \multicolumn{4}{c}{\textit{I believe technology can do a lot of good, but it can also be stressful.}} \\ \\
    Interviewer & 
    \textit{Can you describe in detail what you were doing when you felt stressed?} & \textit{You said that technology can be stressful, can you give me a recent example of when you experienced this?} & \textit{
    You said technology can be stressful, can you clarify what that means to you?} & \textit{Why do you believe that technology can be stressful?}\\
    \bottomrule
\end{tabularx}
\end{table*}

Recent work highlights the limitations of these frameworks~\cite{Robinson2023}, shedding light on several critical areas for improvement. One major critique involves the misuse of silence and encouragement as probes. While traditionally viewed as an effective strategy to elicit responses, silence can create discomfort or confusion, leading to less authentic participant responses. Encouragement probes, intended to motivate further discussion, may inadvertently introduce bias by signalling approval or disapproval. Additionally, Robinson~\cite{Robinson2023} points out the omission of specific probe types like explanatory and idiographic probes in earlier frameworks. Explanatory probes are essential for understanding the reasoning behind participants' actions and thoughts, providing a deeper comprehension of their experiences. Idiographic probes, on the other hand, focus on capturing the unique, individual aspects of participants' experiences, which are often lost in more generalised probing strategies. Robinson~\cite{Robinson2023} advocates for an updated probing framework that includes these overlooked probe types and offers a more comprehensive framework for interviewers. Such a framework, Robinson argues, should emphasise the theoretical underpinnings of each probe type, guiding interviewers to use them more effectively and appropriately in different contexts. 
In light of this, our paper adopts the four theory-based interviewing probes proposed by Robinson~\cite{Robinson2023}. These probes are designed to capture a broad spectrum of participant experiences without leading the conversation. By grounding our probing strategy in established theory, we aim to enhance the quality and reliability of our findings.
Table~\ref{tab:hciexamples} provides examples of each of these four probes within an HCI context. We next describe each of these four probes in detail illustrating their theoretical foundations.

\subsubsection*{Descriptive Probe}
Descriptive probes derive from the Narrative Theory, which posits that individuals recount their lives through episodic narratives comprising both external actions and internal consciousness~\cite{bruner1990acts}. This theory underscores the co-construction of narratives, a concept introduced by Gorden~\cite{gorden1987interviewing} and expanded upon by Pasupathi~\cite{Pasupathi2001}, wherein narratives emerge collaboratively between the interviewer and interviewee. The descriptive probe encapsulates inquiring about both an individual's emotions, thoughts, and motivations during a recounted event as well as details about the surrounding circumstances, activities, and other people involved.

\subsubsection*{Idiographic Probe}
The idiographic probe, grounded in Autobiographical Memory Theory~\cite{Brewer1986}, is an interviewing technique used to elicit detailed and specific memories as opposed to generic and broad recollections. Autobiographical memory stores both types of memories: generic memories that summarise repeated experiences, and specific memories that contain rich, episodic details. Specific memories offer details and contexts, as stored in narrative-like structures that lend themselves to story-telling~\cite{Conway2005, thomsen2009}. Idiographic probing is a method that shifts an interviewee's recall from general to specific by requesting them to describe a single, detailed event that exemplifies a generic memory~\cite{Rottenberg2006}. For example, an idiographic example-type probe might ask for a particular instance that reflects a frequently mentioned experience, thereby facilitating the cognitive transition into specific memory recollection~\cite{thomsen2009, Smith2003}. 

\subsubsection*{Clarifying Probe}
The clarifying probe is a technique informed by Self-Disclosure Theory, which suggests that individuals share personal information in a layered fashion, contingent upon trust and intimacy levels with the listener~\cite{littlejohn2010theories}. As self-disclosure progresses, both the breadth and depth of the information shared increase, with breadth referring to the variety of topics and depth to the degree of detail and time spent on these topics~\cite{west2010introducing, Jourard1971}. Clarifying probes play an important role in interviews, as they are designed to unpack earlier statements to reveal deeper, implicit meanings. By asking interviewees to expound upon a word or phrase, clarifying probes encourage the articulation of thoughts and emotions that might otherwise remain unvoiced.

\subsubsection*{Explanatory Probe}
Explanatory probes are informed by Attribution Theory, which explores how individuals discern the causes of events in their lives. The theory proposes that people attribute causes to events based on a range of factors~\cite{kassin2023social}. These attributions are subjective and influenced by cognitive biases, yet they significantly impact emotions and behaviours. Therefore, understanding the perceived causes behind actions and events is crucial to grasp the narrative sense-making processes that individuals employ. Explanatory probes solicit personal insights into the reasons behind occurrences or feelings. Researchers can use these probes to uncover narrative attributions that participants make regarding the occurrence of certain phenomena~\cite{tilly2006happens}. These probes are particularly useful when the understanding of perceived causality is of interest and can help make sense of autobiographical stories.

\section{Survey Chatbot}
We designed and implemented a chatbot with the goal of 
collecting in-depth qualitative data in online surveys. Specifically, we investigate four distinct interviewing probes~\cite{Robinson2023}, as detailed in Section~\ref{sec:probes}, in the context of three stages of HCI inquiry~\cite[pp. 180-187]{lazar_research_2017}. We implemented the chatbot as a stand-alone application developed in Nextjs. The chatbot is running on a university server and integrates into already established survey platforms as an HTML iframe element~\cite{w3schools_iframe}. This allows for the survey platform to save the data directly alongside the other collected survey data. 

\begin{figure*}[b]
    \centering
    \includegraphics[width=0.56\linewidth]{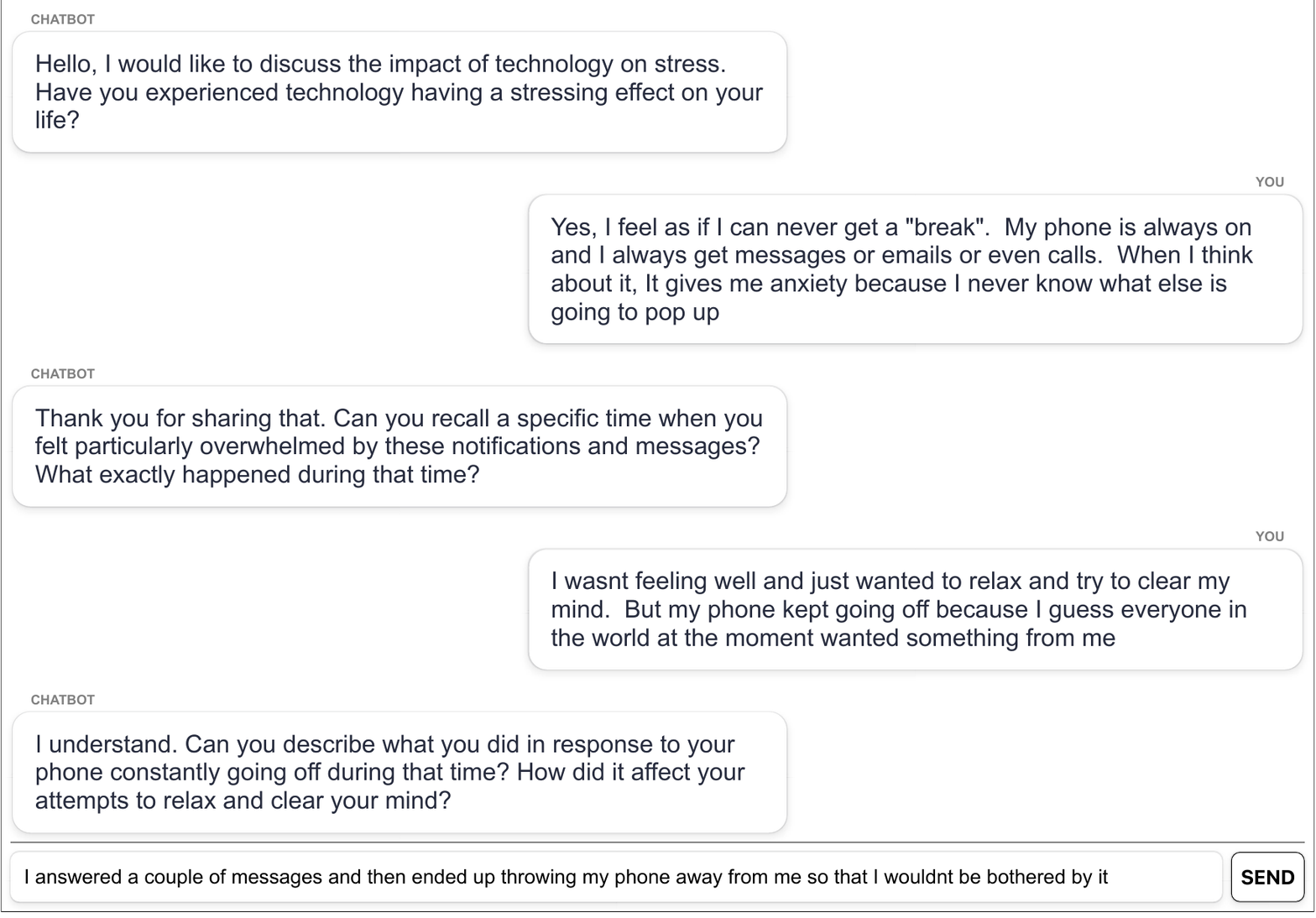}
    \caption{Chatbot interface with a conversation from our study in exploration stage with the idiographic interview probe.}
    \Description{The interface of the chatbot with a conversation from our study in exploration stage with the idiographic interview probe.}
    \label{fig:enter-label}
\end{figure*}

\subsubsection*{Chat Interface}
We have purposefully designed the chat interface to follow simple and minimalistic design considerations (see Fig.~\ref{fig:enter-label}), as both colours~\cite{Hassib2017, Chen2021, Griggio2021, An2024} and the introduction of avatars~\cite{Kurniawan2015, Cortes2023, fota2022, Wonchul2024,liao2020racial} can influence the user experience and perception of a chat conversation. The design follows a conventional text-based instant messaging approach, with a list of message bubbles and an input area at the button with a submit button. The label above each message signifies the sender, in our case: \textit{chatbot} and \textit{you}. We implemented this tool from scratch to enable customisations dedicated to specific research needs, avoid potential limitations of existing libraries, and allow potential future adopters to more easily grasp and control the system's functionality.

\subsubsection*{LLM \& Prompt Architecture}
The chatbot is based on OpenAI's gpt4o model (gpt-4o-2024-05-13) which is instructed through elaborate prompt engineering. The prompt architecture for the chatbot is divided into four discrete parts:

\begin{enumerate}
    \item \textbf{General Instructions} --- This prompt instructs a chatbot to conduct interviews on technology-induced stress, asking one concise question at a time and acknowledging responses neutrally.
    \item \textbf{Topic of Interest} --- This prompt provides a comprehensive summary of technologically induced stress (elaborated in Section~\ref{stress_case}). The goal is to provide the chatbot with a reference for establishing the follow-up questions. 
    \item \textbf{Interview Stage} --- This prompt describes the interview stages of HCI research (elaborated in Section~\ref{applications}). By explaining through the prompt the motivations and goals of each stage the chatbot may adapt the questions to match the current stage. The prompt is being switched according to which stage the chatbot is inquiring into (\ie exploration, requirements gathering, or evaluation).
    \item \textbf{Interview Probe} --- Finally, we provide the bot with a prompt describing the specific interview prompt in detail, including example questions. This prompt instructs the chatbot to follow a specific procedure when inquiring in greater detail based on what participants have already disclosed. This prompt is likewise switched around. 
\end{enumerate}




\section{Method}
We set out to systematically investigate the effects of four distinct interviewing probes on participant responses in conversational surveys.
We investigate this effect across different stages of interviews, reflecting distinct interview goals within HCI.
We structured the study using a 4 (interview probes) $\times$ 3 (stages of interview) mixed design (\ie split-plot design). 
This design allows us to assess the impact of specific probes on the distinct interview stages, reflecting contemporary real-world HCI practice (RQ1). Our split-plot design ensures that participants experience the same interview probe over different interview stages, providing insights across stages while limiting participant load and fatigue. 
We manipulated the interview stage as a within-subjects variable, with each participant experiencing all three interview stages. Conversely, we treat the interview probe as a between-subjects variable, with each participant interacting with only one out of the four interview probes throughout the entire study. All participants focused on the topic of stress related to technology use for the survey. Additionally, we investigate 
the user experience of interacting with an LLM-based chatbot to answer qualitative survey questions through standardised and custom Likert-scales (RQ2).

\subsection{Interview Stages}\label{applications}
Interviews are employed to aid researchers in establishing an understanding of the \iQuote{needs, practices, concerns, preferences, and attitudes} of users towards a current or future technology system~\cite{lazar_research_2017}. We follow the distinctions of three applications of interviews within HCI by Lazar et al.: \texploration, \trequirements, and \tevaluation~\cite[pp. 180-187]{lazar_research_2017}. For clarity, we use the term interview stages in the context of this paper. \texploration~focuses on understanding users' experiences, needs, and preferences involving current or future technologies. This method gathers insights into users' daily practices, aspirations, and challenges with current tools. \trequirements~involves interviewing users to understand their needs, goals, and frustrations within the context of interest with the goal of establishing a set of requirements for future technologies or changes in existing technologies. \tevaluation~is used to gather feedback on prototypes and completed products, crucial for refining design and functionality. \tevaluation~interviews help validate design decisions and identify areas for improvement by capturing user feedback. While we utilise three distinct stages of interviews within HCI, we recognise that the different activities can potentially overlap and often occur in an iterative process. 

\subsection{Technology Stress as a Case}\label{stress_case}
For this particular study, we investigate the case of stress related to, or induced by, the adoption and use of technology~\cite{Agogo2018, Ayyagari2011}. In HCI, this phenomenon (also referred to as \textit{technostress}~\cite{Kupersmith1992, Riedl2012, Dragano2020}, and \textit{technophobia}~\cite{Khasawneh2018, Grudin2005}) has emerged as a significant area of concern~\cite{Cecchinato2020, Fleck2015}. Technology stress concerns the stress or psychological discomfort experienced by individuals due to their inability to adapt or cope with information and communication technologies in a healthy manner. This condition is triggered by the pervasive integration of digital technologies into both our personal and professional lives~\cite{Mark2014}, potentially leading to a state of connectivity that individuals find challenging to manage effectively. Both the absence and presence of technology can be a stressor in people's everyday and working lives~\cite{Notara2021}. The case represents a complex and pertinent area of inquiry holding both apparent and tacit participant knowledge with both explicit and implicit dimensions, making it suitable for investigation in our study.

In our study, all participants went through all three stages of interviews in HCI (\texploration, \trequirements, and \tevaluation). We presented an introduction to the specific task alongside a visual aid (see Appendix~\ref{appen_tasks} for task descriptions and images/video), followed by participants interacting with the chatbot as per one of the four interviewing probes. In \texploration, we showed participants images of everyday activities regarding technology, spanning pictures from private, public, and work contexts in order to provide relatable examples for participants to reflect on their own experiences. In \trequirements, we likewise showed the participants pictures of everyday situations however this time the pictures was with focus on technology in use. Furthermore, the text explicitly stated the goal of setting requirements for interventions into existing or future use of technology. In \tevaluation, the participants watched an introduction video for Apple's `\textit{do not disturb}'-feature for an iOS-based device. In this case, the goal was to elicit particular reflections on a technology's ability to support a user's technologically induced stress. After the preparation step, the user interacted with the chatbot. For each step, the participant was instructed to continue the conversation as long as they had something they were willing to share on the matter. As there are no clear conventions on how to terminate conversations with a chatbot, we instructed participants to send a message saying `goodbye' to the chatbot in order to proceed to the next part of the survey.

\subsection{Participants}
Participants were recruited through Prolific, an online crowdsourcing market for research. 
We restricted participation to those in the participant pool with an acceptance rate above 95\%, based in the USA and with the primary language of English. We limited participation to participants using a desktop computer, as extended typing on a mobile phone can be tiresome and to avoid potential usability issues with the chatbot on a small screen\footnote{In Prolific, we indicated that the study should be taken on desktops only: \url{https://researcher-help.prolific.com/en/article/9c108d}, and we further restricted the participation through Qualtrics: \url{https://www.qualtrics.com/support/survey-platform/survey-module/question-options/display-logic/}.}. We compensated participants with a fixed amount of \$3 for an estimated completion time of 15 minutes. To minimise type II errors, we determined our sample size based on a power calculation using G*Power~\cite{Faul2009}. Considering the exploratory nature of our study, we followed established methodological guidelines~\cite{hair2010multivariate}, using medium-to-large effect sizes \(f^2 = 0.2\), an alpha level of 0.05, and a power of 0.8. This resulted in a minimum sample of 64 participants (16 in each split-plot). We set up four different Prolific studies (\ie one for each interview probe), with a restriction ensuring that a participant can only participate once across all four studies. As such, each group were randomly sampled from the Prolific participant pool matching the restrictions. After accepting the study, the participants were then routed to Qualtrics for survey completion. 

\subsection{Measures}
We collect three sets of measures for our study. First, we evaluate the quality of participant responses both computationally and with manual labelling. Second, we measure participants' experiences and perceptions of the chatbot through both standardised and custom Likert-scales. Finally, we ask participants open-ended questions about the chatbot's way of asking questions and their preferences for providing qualitative data in online surveys. 

\begin{table*}[b]
\caption{Gricean Maxim for analysing conversation quality in conversational surveys, as based on Xiao et al.~\cite{Xiao2020}.}
\label{tab:gricean}
\begin{tabularx}{\linewidth}{@{}ABCD@{}}
\toprule
\textbf{Maxim} & \textbf{Definition} & \textbf{Quality Metric} & \textbf{Definition}
    \\ \midrule
    \textbf{Quantity} & One should be as informative as possible & \textit{Informativeness} & A participant’s response should be as informative as possible \\
    & & \textit{Specificity} & A response should be as detailed as possible \\
    \textbf{Relevance} & One should provide relevant information & \textit{Relevance} & A participant’s response should be relevant to a question asked \\
    \textbf{Manner} & One should communicate in a clear and orderly manner & \textit{Clarity} & A participant’s response should be clear \\
    \bottomrule
\end{tabularx} 
\end{table*}

\subsubsection*{Response Quality}
To assess the quality of the participants' responses we build on the work of ~\citet{Xiao2020}, which introduced the concept of Gricean Maxims to measure the informativeness, relevance, specificity, and clarity of participant responses. Proposed by H.P. Grice in 1975~\cite{Grice1975, grice1989}, Gricean Maxims are described as \iQuote{cooperative principles to guide effective communications}~\cite{dybkjaer1996}. We adopt the Gricean Maxim implementation as proposed by~\citet{Xiao2020} to measure and calculate the participants' response quality. They propose to create proxies for each quality measure which allow for the quantification of each principle, which are defined as \textit{informativeness}, \textit{specificity}, \textit{relevance} and \textit{clarity}. Unlike~\citet{Xiao2020},  we considered every response as we are not only interested in the combined responses, but each response individually. We consider each message as a response to a probing question. This is why we filter out the first response from each condition, as they are not the direct results of a probe. We did not only analyse the individual responses but also each Gricean Maxim in relation to the questions asked and the context. Hence, aiming to provide a focused analysis of the impact of each interview probe. For instance, regarding relevance, a question from the chatbot may be phrased as a ``how do you think question'', but it is answered by listing stressing technologies without providing an explanation -- hence focusing on the \textit{what} rather than the \textit{how} leading to a response of lower relevance.

According to the Gricean Maxim of quantity, effective communication should be informative. We calculated the \textit{informativeness} of text responses~\cite{jones1979elementary} using the sum of each word's surprisal, defined as the inverse of its frequency in modern English. Lower-frequency words convey more information. We averaged word frequencies across four text corpora---the British National Corpus~\cite{Leech_1993}, The Brown Corpus~\cite{hofland1982}, Web Text Corpus~\cite{parviainen2010}, and the NPS Chat Corpus~\cite{Forsythand2007}, again following Xiao et al.~\cite{Xiao2020})---to estimate a word's commonness in modern English and subsequently assessed the total informativeness of participants' responses to open-ended questions based on this metric. 

While our informativeness metric quantifies the amount of information in a user’s text response, it does not evaluate the \textit{specificity} of the response. Specific responses, which provide detailed information, aid researchers in understanding and leveraging the data, and offer deeper insights. Given the variability and complexity of responses to open-ended questions, we manually rated each response’s specificity on a three-point scale: 0 for generic descriptions, 1 for specific concepts, and 2 for specific concepts with detailed examples. 

According to the Gricean Maxim of \textit{relevance}, quality communication must be pertinent to the context. In surveys, responses should directly address the posed question. Irrelevant responses not only lack value but also complicate analysis. Given the diversity and complexity of responses to open-ended questions, we manually evaluated their relevance on a three-point scale: 0 for irrelevant, 1 for somewhat relevant, and 2 for fully relevant. 

Finally, according to the Gricean Maxim of \textit{clarity}, effective communication should be clear and easily understood without ambiguity. We assessed the clarity of each text response on a three-level scale: 0 for illegible text, 1 for incomplete sentences or grammatical errors that hinder understanding, and 2 for clearly articulated responses with complete sentences and no significant grammatical issues.

For the metrics, \textit{specificity}, \textit{relevance} and \textit{clarity}, we provide examples in Appendix~\ref{example_responses} on the scoring of text responses on the three-point scale. Our labelling process involved two researchers initially coded a random 10\% of the dataset blind to which interview probe was used. The two authors were intentionally aware of the research stage to be able to evaluate whether a response was considered of good quality and in line with the goals in the research stage. Following this, the two authors met and compared the labelled responses and resolved any differences by critically evaluating the criteria and arguments for every label of a response. After every response label was settled, the first author labelled the remainder of the dataset. 

\subsubsection*{Subjective Experience}\label{sec:subj_exp}
We measure the participants' subjective experience of interacting with the chatbot by asking them to answer a set of Likert-scale questions (see Table~\ref{tab:subjective_measures}). First, we utilise the \textit{smoothness}-scale of the standardised Session Evaluation Questionnaire (SEQ)~\cite{Stiles1994}, which is scored on five Likert-style questions ranging from 1--7 (see Appendix~\ref{seq_calc} for dimension calculations, and~\cite{Stiles} for instructions). In recent work, SEQ has been used to measure the experiences of users interacting with chatbots in HCI~\cite{bell2019}. Second, we ask the participant to rate the chatbot on four Likert-point scales whether they experienced the chatbot as \textit{helpful}~\cite{zarouali2018predicting}, \textit{useful}~\cite{zarouali2018predicting}, \textit{repetitive}, and \textit{intrusive}. Finally, we ask the participants to rate on a 7-point Likert-scale if the things they revealed to the chatbot during the three sessions are accurate reflections on their personal thoughts, feelings, and experiences~\cite{kang2024counseling,wheeless1976self}.

\begin{table*}[t]
\caption{Overview of the Likert-scales for measuring subjective experience}
\label{tab:subjective_measures}
\begin{tabularx}{\linewidth}{@{}HIJK@{}}
\toprule
\textbf{Measure} & \textbf{Question} & \textbf{Scale} & \textbf{Ref.} 
    \\ \midrule
    SEQ & My conversations with the chatbot were...: <Difficult-Easy>, <Relaxed-Tense>, <Unpleasant-Pleasant>, <Rough-Smooth>, and <Comfortable-Uncomfortable> & 1--7 & \cite{Stiles1994} \\ 
    Helpful & As an interviewee, I perceived the chatbot's questions as: Helpful <Strongly Disagree-Strongly Agree> & 1--7 & \cite{zarouali2018predicting} \\
    Useful & As an interviewee, I perceived the chatbot's questions as: Useful <Strongly Disagree-Strongly Agree> & 1--7 & \cite{zarouali2018predicting} \\
    Repetitive & As an interviewee, I perceived the chatbot's questions as: Repetitive <Strongly Disagree-Strongly Agree> & 1--7 & \\
    Intrusive & As an interviewee, I perceived the chatbot's questions as: Intrusive <Strongly Disagree-Strongly Agree> & 1--7 & \\
    Accuracy & The things that I revealed to the chatbot are accurate reflections of my personal thoughts, feelings and experiences <Strongly Disagree-Strongly Agree> & 1--7 & \cite{kang2024counseling,wheeless1976self} \\
    \bottomrule
\end{tabularx} 

\end{table*}

\subsubsection*{Post-study Open-ended Questions}
At the end study, we asked the participants four open ended questions in two parts. First, to better understand the participants' experiences and perceptions of the chatbot itself, we asked them to describe their experiences with the chatbot's line of questioning and the chatbot's questioning style. Second, to understand the use of chatbots as part of surveys, we asked the participants to compare their experience of responding to questions from a chatbot compared to traditional open-ended questions, thereby drawing from the participants' vast experience as frequent survey participants. We also asked the participants whether there are any topics that they would rather disclose to a chatbot than a human, or vice versa. We asked the following questions:

\begin{itemize}[topsep=0pt]
    \item Please describe your experience with and perceptions of the chatbot's line of questioning. For example how the chatbot's questions affected your answers. 
    \item How would you describe your perceptions of the chatbot’s questioning style? Please consider the variety, relevance, and impact of the questions asked.
    \item How does your experience of talking to the chatbot compare to responding to traditional open-ended surveys? Please describe specific drawbacks or benefits.
    \item Are there any topics that you would rather disclose to a chatbot than to a human, or vice versa? Please motivate why / why not.
\end{itemize}

\subsection{Data Analysis}
\subsubsection*{Response Quality}
To investigate the effect of our independent variables (interview probe and interview stage) on the quality measures of \textit{informativeness}, \textit{relevance}, \textit{specificity}, and \textit{clarity}, we constructed a mixed model for each of the four quality measures. We conducted an iterative construction of each of the models through incremental removal of variables based on their predictive power and Akaike information criterion (AIC), including number of messages, age, and gender of the participants, as well as duration of interaction. Our final set of predictors consisted of the two independent variables, interview probes and interview stages, for all models. As we include each individual response in our models, we included participant ID as a random effect in all four models to account for variance between participants, mitigating inflated Type I errors due to non-independence of observations. The informativeness model was constructed and evaluated using the R package lme4~\cite{lme4}, and the relevance, specificity, and clarity models were constructed as Cumulative Link Mixed Models due to their ordered ordinal data.

\subsubsection*{Subjective Experience}
To analyse the multivariate Likert data (chatbot experience, smoothness, and accuracy), we employed the Aligned Rank Transform (ART) method as proposed by Wobbrock et al.~\cite{wobbrock2011}, using the ARTool R package. ART is well-suited for non-parametric data, while preserving the assumptions of traditional parametric tests. We investigated the effects of the independent variable \textit{interview probe}. Subsequently, we performed pairwise \textit{t}-tests to assess the significance of these variables, applying Bonferroni correction to adjust for multiple comparisons.

\subsubsection*{Analysis of Post-study Open-ended Questions}
We grouped the answers to the four open-ended questions into two categories: one for experiences and perceptions of questioning style, and the other for opinions on chatbots in surveys. 
We then looked for the most salient patterns in the answers of each group following an inductive approach~\cite[pp. 285-299]{lazar_research_2017}. The first author annotated the data with an initial set of codes in a spreadsheet, which was then refined in discussion with another author.  Continuing this, the first author wrote a summary of the most salient patterns, which were further discussed and iterated with the rest of the authors.

\subsubsection*{Exploratory Conversation Analysis}
As part of our investigation into open-ended textual conversations between survey participants and a chatbot, we sought to assess emerging conversational patterns. The first author reviewed the conversation history for each participant, \change{identifying communication and interaction patterns} which were then discussed with the co-authors, with the most significant interactions selected and exemplified. Through iteration, we selected a final set of five distinct patterns. This exploratory analysis was not intended to generate generalisable claims, but rather to highlight unique aspects of the conversations that could deepen our understanding of participant-chatbot interactions and inspire future designs.



\section{Results}
We report the results of our study, which compared four theory-based interview probes across three stages of HCI research. A total of 64 participants (16 per split-plot), evenly divided by gender (32 male, 32 female) and ranging in age from 19 to 67 years ($M~=38.67$, $SD=12.31$), took part in the study. The participants generated 1,287 responses to the interview probe questions across all conditions (see Table~\ref{tab:messages_sent} for the distributions). In this section, we first present the findings related to quality measures across all conditions. Next, we provide insights into participants' perceptions and experiences of interacting with the chatbot. Third, we explore qualitative feedback regarding the chatbot’s questioning style and its potential use in surveys. Finally, we offer exploratory insights into interesting patterns in the interactions participants had with the chatbot. 

\begin{table}[t]
\caption{Messages sent across conditions.}
\label{tab:messages_sent}
\resizebox{\linewidth}{!}
{%
\begin{tabular}{@{}llll@{}}
\toprule
& \textbf{Exploration} & \textbf{Requirements} & \textbf{Evaluation} \\ 
\midrule
\textbf{Descriptive} & 114 ($M~=~8.14$) & 115 ($M~=~7.19$) & 88 ($M~=~7.19$) \\
\textbf{Idiographic} & 115 ($M~=~7.67$) & 69 ($M~=~4.6$) & 44 ($M~=~2.93$) \\
\textbf{Clarifying} & 162 ($M~=~10.1$) & 97 ($M~=~6.93$) & 79 ($M~=~5.27$) \\
\textbf{Explanatory} & 190 ($M~=~11.9$) & 141 ($M~=~8.81$) & 73 ($M~=~5.62$) \\
\bottomrule
\end{tabular}
}
\end{table}

\begin{table*}[b]
 \caption{Generalised linear models of the four quality measures. For each predictor, we report coefficients, standard errors (in brackets), and significance indicators. The sign of the estimate (+/-) denotes the direction of the relationship between the predictor and outcome variable (informativeness, relevance, specificity, and clarity).}
   \label{tab:quality_model}
\resizebox{\textwidth}{!}{%
  \begin{tabular}{@{}lrlrlrlrl@{}}
    \toprule
    \textbf{Predictor} & \multicolumn{2}{c}{\textbf{Informativeness}} & \multicolumn{2}{c}{\textbf{Relevance}} & \multicolumn{2}{c}{\textbf{Specificity}} & \multicolumn{2}{c}{\textbf{Clarity}}\\
    \midrule
    \multicolumn{5}{l}{\textit{\textbf{Baselines:}}}\\ 
    \multicolumn{5}{l}{\textit{\textbf{Probe = Descriptive, Interview Stage = Exploration}}} \\\\ 
        \textbf{Interview Probe} \\
    \hspace{3mm}Idiographic                             & 32.95 & (52.02) & 0.24 & (0.48)       & 0.21 & (0.21) & 0.23 & (0.48) \\
    \hspace{3mm}Clarifying                              & -65.90 & (51.37) & -0.99 & (0.48)*    & -0.20 & (0.43) & -0.99 & (0.46) \\
    \hspace{3mm}Explanatory                             & -71.62 & (50.87) & -1.46 & (0.46)**   & -1.30 & (0.43)** & -1.46 & (0.46)** \\
    \textbf{Interview Stage} \\
    \hspace{3mm}Requirements                            & -14.10 & (-14.10) & -0.36 & (0.29)    & -0.70 & (0.28)* & -0.36 & (0.29) \\
    \hspace{3mm}Evaluation                              & -64.02 & (-64.02)** & -1.04 & (0.29)*** & -1.19 & (0.28)*** & -1.03 & (0.30) \\
    \textbf{Interaction effects} \\
    \hspace{3mm}Probe (Idiographic) : Stage (Requirements)& -15.29 & (-15.29) & 1.07 & (0.44)*    & 0.24 & (0.36) & 1.06 & (0.44)* \\
    \hspace{3mm}Probe (Clarifying) : Stage (Requirements) & 5.58 & (5.58) & 1.08 & (0.41)**       & -0.31 & (0.38) & 1.07 & (0.40)** \\
    \hspace{3mm}Probe (Explanatory) : Stage (Requirements)& -52.75 & (-52.75) & 0.08 & (0.38)    & -0.11 & (0.36) & 0.08 & (0.38) \\
    \hspace{3mm}Probe (Idiographic) : Stage (Evaluation)  & 35.58 & (35.58) & 1.54 & (0.49)**    & 1.08 & (0.46)* & 1.54 & (0.50)** \\
    \hspace{3mm}Probe (Clarifying) : Stage (Evaluation)   & 22.57 & (22.57) & 1.61 & (0.42)***   & -0.07 & (0.40) & 1.61 & (0.42)*** \\
    \hspace{3mm}Probe (Explanatory) : Stage (Evaluation)  & -8.39 & (-8.39) & 1.32 & (0.42)**    & 0.87 & (0.40)* & 1.32 & (0.42)** \\
  \bottomrule
  \end{tabular}
  }
\begin{flushleft} 
  *{***\textit{p} < 0.001, **\textit{p} < 0.01, *\textit{p} < 0.05}
\end{flushleft}
 \Description{Generalised linear models of the four quality measures. For each predictor, we report coefficients, standard errors (in brackets), and significance indicators. The sign of the estimate (+/-) denotes the direction of the relationship between the predictor and outcome variable (informativeness, relevance, specificity, and clarity).}
\end{table*}

\subsection{Assessing Response Quality}
We used a likelihood ratio test for comparing each model to its respective null model, which shows that our logistic regression models are all statistically significant; informativeness ($\chi^2$(11) = 40.412, \textit{p} < .001), relevance ($\chi^2$(11) = 48.928, \textit{p} < .001), specificity ($\chi^2$(11) = 71.035, \textit{p} < .001), and clarity ($\chi^2$(11) = 86.451, \textit{p} < .001). The informativeness model explains 45.8\% of the variance in our model ($\mathbb{R}$ = 0.68, $\mathbb{R}^{2}$ = 0.46), the relevance model 28.2\% ($\mathbb{R}$ = 0.53, $\mathbb{R}^{2}$ = 0.28), the specificity model 28.4\% ($\mathbb{R}$ = 0.53, $\mathbb{R}^{2}$ = 0.28), and the clarity model 33.9\% ($\mathbb{R}$ = 0.58, $\mathbb{R}^{2}$ = 0.34). We tested for the existence of multicollinearity among the models' parameters and found a variation inflation factor (VIF) between 1.65 and 5.11 for the predictors of the informativeness model, between 1.89 and 5.57 for the relevance model, between 1.87 and 5.09 for the specificity model, and between 1.80 and 5.10 for the clarity model. These values are all below the often-used threshold of ten to detect multicollinearity~\cite{hair2010multivariate}.
We now present the significant predictors of all four models. Outcomes for all models are summarised in Table~\ref{tab:quality_model}. 

\subsubsection*{Interview Probe}
In this section, we present the main effects of interview probe type on the four quality measures (see Figure~\ref{fig:probe_main_effects}). For relevance, we find that the \texplanatory~probe provides significantly less relevance compared to the \tidiographic~probe (\( \hat{\beta} = -2.1000 \), \textit{SE} = 0.425, \textit{p} < 0.0001) and the \tclarifying~probe (\( \hat{\beta} = -1.1972 \), \textit{SE} = 0.429, \textit{p} = 0.0270). In terms of specificity, the \texplanatory~probe elicited significantly less specific responses compared to the \tdescriptive~probe (\( \hat{\beta} = 1.053 \), \textit{SE} = 0.391, \textit{p} = 0.0358) and the \tidiographic~probe (\( \hat{\beta} = -1.714 \), \textit{SE} = 0.397, \textit{p} = 0.0001). Regarding clarity, both the \tclarifying~and \texplanatory~probes result in notably lower scores as compared to the \tdescriptive~and \tidiographic~probes, suggesting that these two probe types might be less effective in prompting clear responses. Regarding clarity, the \texplanatory~probe resulted in significantly lower scores compared to both the \tdescriptive~probe (\( \hat{\beta} = 1.619 \), \textit{SE} = 0.515, \textit{p} = 0.0092) and the \tidiographic~probe (\( \hat{\beta} = -2.738 \), \textit{SE} = 0.524, \textit{p} < 0.0001). The \tclarifying~probe also scored significantly lower than the \tidiographic~probe (\( \hat{\beta} = -1.724 \), \textit{SE} = 0.524, \textit{p} = 0.0055). Finally, for informativeness, we find no significant differences between the four probes.

\setcounter{figure}{0}
\begin{figure*}[t]
    \centering
    \includegraphics[width=1\linewidth]{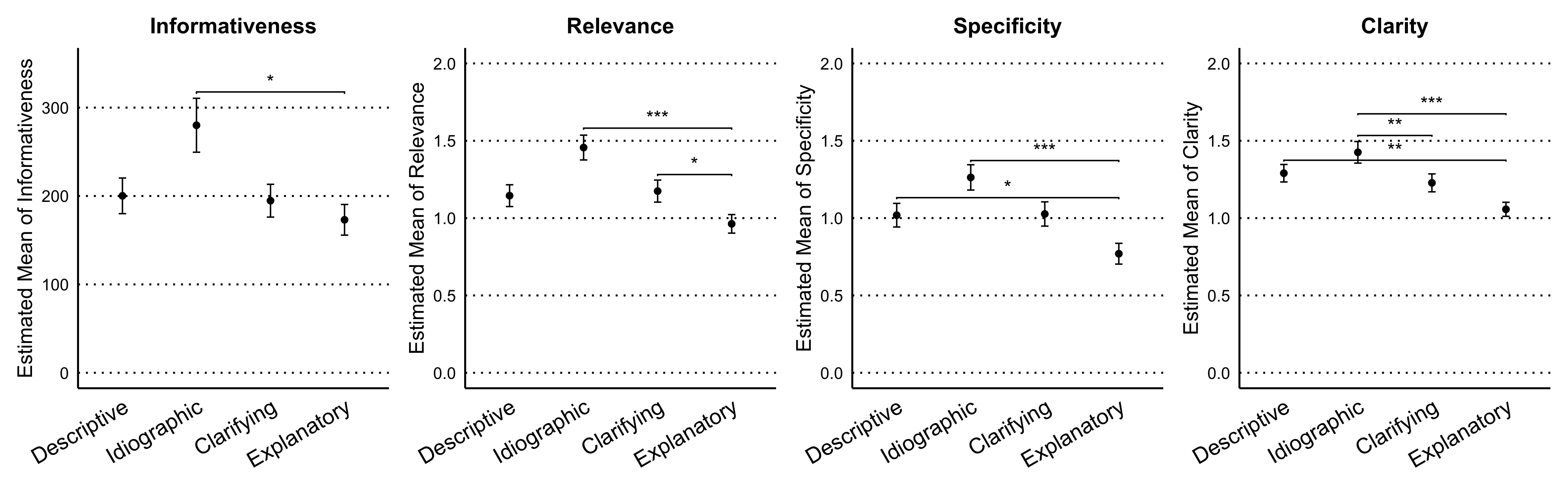}
    \caption{Visualisation of the interview probe main effects for each quality measure.}
    \Description{Visualisation of the interview probe main effects for each quality measure.}
    \label{fig:probe_main_effects}
\end{figure*}

\begin{figure*}[b]
    \centering
    \includegraphics[width=1\linewidth]{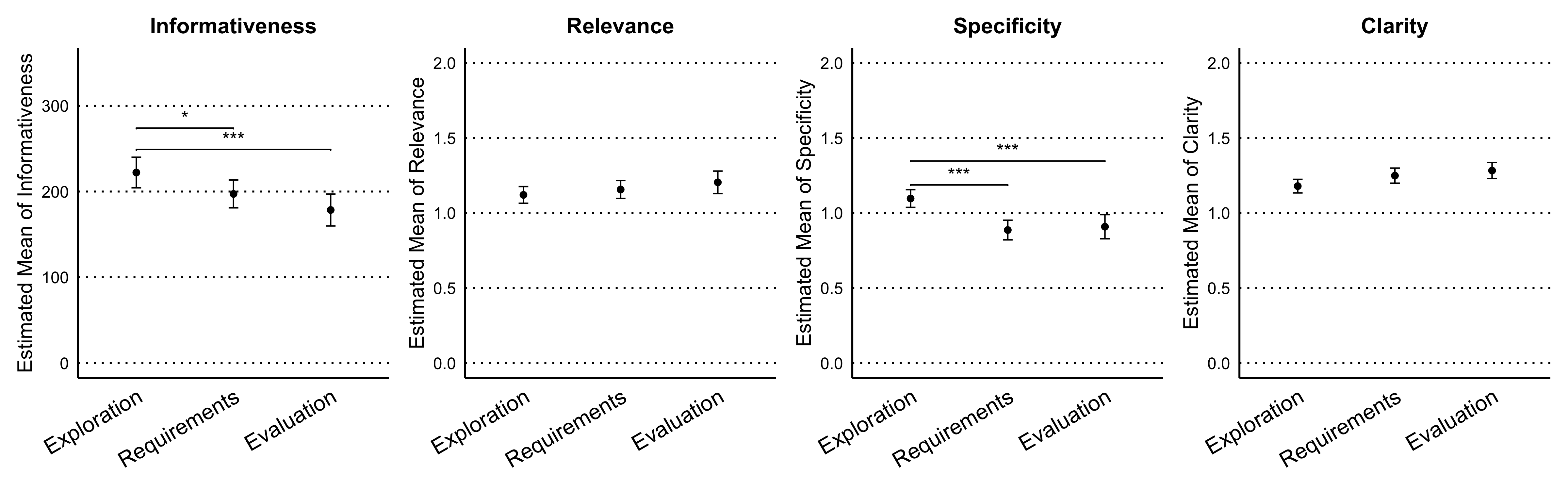}
    \caption{Visualisation of the interview stage main effects for each quality measure.}
    \Description{Visualisation of the interview stage main effects for each quality measure.}
    \label{fig:main_application_effects}
\end{figure*}

\subsubsection*{Interview Stage}
Next, we present the main effects of interview stage type on the four quality measures (see Figure~\ref{fig:main_application_effects}). For informativeness, the \texploration stage scored significantly higher than both the \tevaluation~(\( \hat{\beta} = 51.6 \), \textit{SE} = 12.0, \textit{p} = 0.0001) and \trequirements~stages (\( \hat{\beta} = 29.7 \), \textit{SE} = 10.7, \textit{p} = 0.0154), indicating that responses in the \texploration~stage were more informative compared to the other stages. The \tevaluation~stage had a significant negative effect on relevance compared to the \texploration~stage (\textit{p} < 0.001). However, post-hoc comparisons did not find significant differences between the stages, suggesting that the stage effect alone may not be strong. For specificity, both the \trequirements~(\( \hat{\beta} = 0.7448 \), \textit{SE} = 0.137, \textit{p} < 0.0001) and \tevaluation~stages (\( \hat{\beta} = 0.7215 \), \textit{SE} = 0.155, \textit{p} < 0.0001) produced significantly less specific responses compared to the \texploration~stage. Finally, no significant differences were observed between the three interview stages in terms of clarity, suggesting that the clarity of responses remained consistent across all stages.

\subsubsection*{Interaction between Probe and Stage}
\begin{figure*}[t]
    \centering
    \includegraphics[width=1\linewidth]{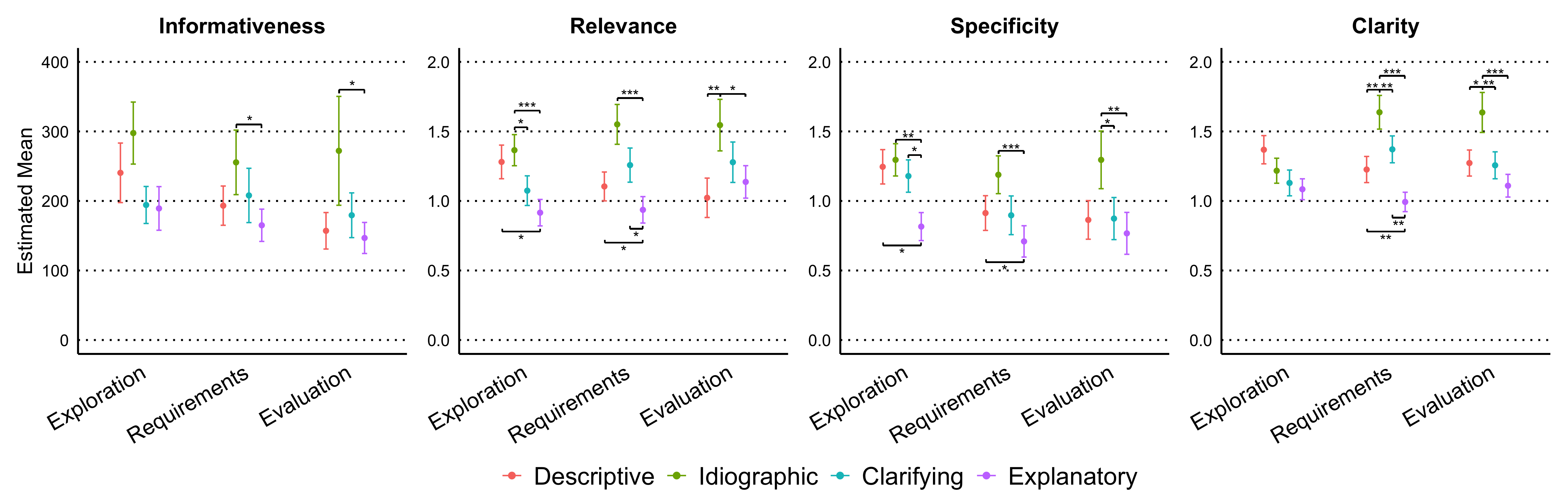}
    \caption{Visualisation of the interaction effects between interview stage and interview probe main for each quality measure.}
    \Description{Visualisation of the interaction effects between interview stage and interview probe main for each quality measure.}
    \label{fig:interaction_effects}
\end{figure*}

Next, we present the results of the interaction effects between the interview probe and interview stage in order to investigate each probe's usefulness within each stage of HCI research. Specifically, we further analyse each significant predictor through a pairwise comparison with Tukey p-value correction. We report how the individual probes compare to each other based in a particular interview stage. We visualise all interaction effects in Figure~\ref{fig:interaction_effects}. 

\begin{table*}[b]
\caption{Quality measures for the three stages of interviews.}
\label{tab:exploration_phase}
\begin{tabular}{@{}llllll@{}}
\toprule
\textbf{Stage} & \textbf{Probe} & \textbf{Informativeness} & \textbf{Relevance} & \textbf{Specificity} & \textbf{Clarity} \\ 
\midrule
\textbf{Exploration} & \textbf{Descriptive} & 240 ($SD = 233$) & 1.28 ($SD = 0.66$) & 1.25 ($SD = 0.67$) & 1.37 ($SD = 0.55$) \\
&\textbf{Idiographic} & 298 ($SD = 244$) &1.07 ($SD = 0.69$) & 1.18 ($SD = 0.76$) & 1.13 ($SD = 0.60$) \\
&\textbf{Clarifying} & 194 ($SD = 173$) & 0.92 ($SD = 0.67$) & 0.82 ($SD = 0.71$) & 1.08 ($SD = 0.53$) \\
&\textbf{Explanatory} & 189 ($SD = 221$) & 1.37 ($SD = 0.61$) & 1.30 ($SD = 0.64$) & 1.22 ($SD = 0.49$) \\ 
\cline{2-6}
\textbf{Requirements} & \textbf{Descriptive} & 193 ($SD = 154$) & 1.10 ($SD = 0.57$) & 0.91 ($SD = 0.68$) & 1.23 ($SD = 0.51$) \\
&\textbf{Idiographic} & 208 ($SD = 196$) & 1.26 ($SD = 0.62$) & 0.90 ($SD = 0.70$) & 1.37 ($SD = 0.49$) \\
&\textbf{Clarifying} & 165 ($SD = 141$) & 0.94 ($SD = 0.58$) & 0.71 ($SD = 0.68$) & 0.99 ($SD = 0.42$) \\
&\textbf{Explanatory} & 255 ($SD = 197$) & 1.55 ($SD = 0.61$) & 1.19 ($SD = 0.58$) & 1.64 ($SD = 0.51$) \\ 
\cline{2-6}
\textbf{Evaluation} & \textbf{Descriptive} & 157 ($SD = 126$) & 1.02 ($SD = 0.68$) & 0.86 ($SD = 0.66$) & 1.27 ($SD = 0.45$) \\
&\textbf{Idiographic} & 179 ($SD = 146$) & 1.28 ($SD = 0.66$) & 0.87 ($SD = 0.69$) & 1.26 ($SD = 0.44$) \\
&\textbf{Clarifying} & 147 ($SD = 97.3$) & 1.14 ($SD = 0.51$) & 0.77 ($SD = 0.66$) & 1.11 ($SD = 0.36$) \\
&\textbf{Explanatory} & 272 ($SD = 265$) & 1.55 ($SD = 0.63$) & 1.30 ($SD = 0.70$) & 1.64 ($SD = 0.49$) \\ 
\bottomrule
\end{tabular}
\end{table*}

\subsubsection*{Interaction between Probe and Stage for Exploration}
We now present the significant results of the pairwise comparisons of interview probes for the \texploration~stage. We provide means and standard deviation for each measure and probe in Table~\ref{tab:exploration_phase}. No significant differences were observed between the probes for informativeness. For relevance, the \tdescriptive~probe scored significantly higher than the \texplanatory~probe (\( \hat{\beta} = 1.4608 \), \textit{SE} = 0.459, \textit{p} = 0.008). The \tidiographic~probe scored significantly higher than the \tclarifying~probe (\( \hat{\beta} = -1.2264 \), \textit{SE} = 0.463, \textit{p} = 0.040), and the \tidiographic~probe also scored significantly higher than the \texplanatory~probe (\( \hat{\beta} = -1.7007 \), \textit{SE} = 0.456, \textit{p} = 0.001). For specificity, the \tdescriptive~probe performed significantly better than the \texplanatory~probe (\( \hat{\beta} = 1.303 \), \textit{SE} = 0.430, \textit{p} = 0.0131), and the \tclarifying~probe also outperformed the \texplanatory~probe (\( \hat{\beta} = 1.099 \), \textit{SE} = 0.414, \textit{p} = 0.0399). Additionally, the \tidiographic~probe performed better than the \texplanatory~probe (\( \hat{\beta} = -1.519 \), \textit{SE} = 0.429, \textit{p} = 0.0023). Finally, we found no significant differences for clarity. 

\subsubsection*{Interaction between Probe and Stage for Requirements Gathering}
We now present the significant results of the pairwise comparisons of interview probes for the \trequirements~stage. We provide means and standard deviation for each measure and probe in Table~\ref{tab:exploration_phase}. For informativeness the \tidiographic~probe outperformed the \texplanatory~probe (\( \hat{\beta} = -142.02 \), \textit{SE} = 52.8, \textit{p} = 0.0421). For relevance, the \tclarifying~probe scored significantly higher than the \texplanatory~probe (\( \hat{\beta} = 1.4717 \), \textit{SE} = 0.474, \textit{p} = 0.0103), and the \tdescriptive~probe also outperformed the \texplanatory~probe (\( \hat{\beta} = 1.3786 \), \textit{SE} = 0.471, \textit{p} = 0.0180). The \tidiographic~probe scored significantly higher than the \texplanatory~probe (\( \hat{\beta} = -2.6851 \), \textit{SE} = 0.504, \textit{p} < 0.0001). For specificity, the \tdescriptive~probe outperformed the \texplanatory~probe (\( \hat{\beta} = 1.419 \), \textit{SE} = 0.448, \textit{p} = 0.0084), and the \tidiographic~probe also performed better than the \texplanatory~probe (\( \hat{\beta} = -1.885 \), \textit{SE} = 0.466, \textit{p} = 0.0003). And finally, for clarity, the \tdescriptive~probe outperformed the \texplanatory~probe (\( \hat{\beta} = 1.3786 \), \textit{SE} = 0.471, \textit{p} = 0.0180), and the \tclarifying~probe also performed better than the \texplanatory~probe (\( \hat{\beta} = 1.4717 \), \textit{SE} = 0.474, \textit{p} = 0.0103). Additionally, the \tidiographic~probe outperformed the \texplanatory~probe (\( \hat{\beta} = -4.0268 \), \textit{SE} = 0.614, \textit{p} < 0.0001), the \tdescriptive~probe (\( \hat{\beta} = -1.9912 \), \textit{SE} = 0.583, \textit{p} = 0.006), and the \tclarifying~probe (\( \hat{\beta} = -1.9135 \), \textit{SE} = 0.612, \textit{p} = 0.001).

\subsubsection*{Interaction between Probe and Stage for Evaluation}
We now present the significant results of the pairwise comparisons of interview probes for the \tevaluation~stage. We provide means and standard deviation each measure and probe in Table~\ref{tab:exploration_phase}. The informativeness score the \tidiographic~probe outperformed the \texplanatory~probe (\( \hat{\beta} = -148.53 \), \textit{SE} = 55.8, \textit{p} = 0.0441). For relevance, the \tidiographic~probe scored significantly higher than the \tdescriptive~probe (\( \hat{\beta} = -1.7753 \), \textit{SE} = 0.558, \textit{p} = 0.0080) and also scored significantly higher than the \texplanatory~probe (\( \hat{\beta} = -1.9142 \), \textit{SE} = 0.559, \textit{p} = 0.0035). For specificity, the \tidiographic~probe outperformed both the \tclarifying~probe (\( \hat{\beta} = -1.577 \), \textit{SE} = 0.521, \textit{p} = 0.0132) and the \texplanatory~probe (\( \hat{\beta} = -1.736 \), \textit{SE} = 0.527, \textit{p} = 0.0055). For clarity, the \tidiographic~probe outperformed the \tdescriptive~probe (\( \hat{\beta} = -1.7753 \), \textit{SE} = 0.558, \textit{p} = 0.0080), the \texplanatory~probe (\( \hat{\beta} = -1.9142 \), \textit{SE} = 0.559, \textit{p} = 0.0035), and the \tclarifying~probe (\( \hat{\beta} = -2.291 \), \textit{SE} = 0.651, \textit{p} = 0.0025).

\subsection{Assessing Participant Perceptions and Experience}

\subsubsection*{Participant Perceptions of the Chatbot} 
We found significant differences between the probes in participants' perceptions of the chatbot's repetitiveness (F(60) = 2.922, \textit{p} = 0.041). Post hoc pairwise comparisons with Bonferroni correction showed significantly higher values for the \texplanatory~probe (\textit{M}~=~6.312, \textit{SD}~=~0.957) compared to the \tclarifying~probe (\textit{M}~=~4.625,\ \textit{SD}~=~1.628), (t(24.268)~=~3.177, \textit{p}~=~0.039, \textit{d}~=~1.123). We found no difference between the two remaining probes, \tdescriptive{} (\textit{M}~=~5.313, \textit{SD}~=~2.056) or \tidiographic~(\textit{M}~=~5.813, \textit{SD}~=~1.109). We found no significant differences between perceived helpfulness (F(60) = 0.733, \textit{p} = 0.536), usefulness (F(60) = 0.600, \textit{p} = 0.617), or intrusiveness (F(60) = 1.473, \textit{p} = 0.231). 

\subsubsection*{Smoothness}
We calculated the smoothness score (see Appendix~\ref{seq_calc}) from the SEQ responses and compared these scores across interview probes. ART showed no significant differences (F(60) = 0.211, \textit{p} = 0.888) between the \tdescriptive{} (\textit{M}~=~5.288,\ \textit{SD}~=~1.592), \tidiographic{} (\textit{M}~=~5.263, \textit{SD}~=~1.047), \tclarifying~(\textit{M}~=~5.025, \textit{SD}~=~1.578), and \texplanatory~(\textit{M}~=~4.913, \textit{SD}~=~1.427) probes.

\subsubsection*{Perceived Accuracy}
We compared the participants' scoring of their perceived ability to be accurate about their opinions and feelings within each condition. The ART analysis showed no significant differences (F(60) = 1.914, \textit{p} = 0.137) between the \tdescriptive{} (\textit{M}~=~6.438, \textit{SD}~=~0.727), \tidiographic{} (M~=~6.750, \textit{SD}~=~0.447), \tclarifying~(\textit{M}~=~6.250, \textit{SD}~=~0.931), and \texplanatory{} (\textit{M}~=~5.938, \textit{SD}~=~1.482) interview probes.

\subsection{Qualitative Insights}
We now provide qualitative insights into the user experiences and perceptions of the chatbot, as well as their perceptions of utilising chatbots as part of surveys. We did not observe a particular pattern between the different probes, which means we will provide the insights on an aggregated basis. 

\subsubsection*{Experiences and Perceptions of Questioning Style}
The open-ended questions regarding experiences and perception of the chatbot's questioning style can be grouped into the ability of the chatbot to facilitate the conversation and disclosure, and the repetitiveness of the probing questions. 

A total of 33 of the participants directly described how the conversational format and the influence of the follow-up questions facilitated them in providing answers to the survey. Participants characterised the chatbot's line of questioning as allowing them to engage in \textit{``deeper''}~(P3, P25, P36) discussions, \textit{``reflect''}~(P3, P19) more before giving an answer, aided in \textit{``narrowing''}~(P59) down their thoughts, providing \textit{``relevant''}~(P15, P24, P25, P28, P47, P53, P55) questions, kept the conversation \textit{``focused''}~(P45) and on \textit{``topic''}~(P28, P39), and generally \textit{``helpful''}~(P2, P17, P38) in helping the participants to express themselves. For instance, participant 57 stated that: \textit{``I do feel that the prompt made me go into more detail than I normally would have gone into. I felt like the questions were open-ended and not leading me in any way''}, indicating that chatbots in surveys may increase the depth of participant responses to researchers' open-ended questions. As illustrated by participant 2: \textit{`The chatbot's questions were helpful in directing my thoughts towards how I felt about experiences, and connecting those feelings with events. The questions were friendly and curious, and actually made me feel seen, and thus more willing to share my thoughts.''}. Other participants stated how they felt better able to; \textit{``elaborate''}~(P24, P27) on their thoughts and feelings, better capable of providing \textit{``detailed''}~(P18, P38, P43, P57, P58) as well as \textit{``meaningful''}~(P3) responses, and that their answering was more \textit{``thorough''}~(P35, P43, P56) than usual. Exemplified by participant 35: \textit{``The chatbot definitely made me much more thorough than I normally would have been with my answers but making sure it understood exactly what I was trying to get across''}.

From the open-ended questions, we observed that 15 participants noted how the chatbot, through longer use, exhibited repetitiveness in its line of questioning. For instance, participant 49 stated that: \textit{``The chatbot asked me questions that seemed a bit repetitive at times. The questions were interesting for a couple of times and then it seemed it asked the same line of questioning after that''}. Highlighting how the lack of an endpoint may lead to experienced repetitiveness of the chatbots' responses. 

\subsubsection*{Chatbots as part of Surveys}
We asked participants both about their opinions on using chatbots as part of surveys in regards to the conventional method of open-ended questions as well as considerations on which topics they would disclose to chatbots and humans. 

Of the 64 participants, 38 would prefer to interact with a chatbot compared to filling out open-ended questions in a survey, ten participants preferred to fill out open-ended questions, and 16 either liked both equally or did not care at all. As stated by participant 2: \textit{``Interacting with the chatbot was better than a traditional open-ended survey. Answering questions is easier, more focused, and more natural than just having to come up with a response. The expectations were clear and the bot adapted well to my style of conversing, so I felt motivated to answer thoroughly and honestly''}. Of the ten participants arguing for open-ended questions, they mentioned the extended time used~(P4, P26, P39, P58), that open-ended questions allow the participants themselves to frame the answers to the question~(P1, P20, P59), and finally, that it can be difficult to know when to stop the conversation~(P46, P59). As illustrated by participant 59 \textit{``They have the benefit of letting me frame my answer the way that I think will communicate my thoughts best. The chatbot does help focus the information, but it can really limit feedback, especially when asking about situations that don't apply.''}.  

When asked if there are any topics they would rather disclose to a chatbot rather than a human in a survey, 14 participants found it easier to disclose to the chatbot, six prefer another human, and 44 did not believe that there are great differences in what they would disclose to a chatbot compared to a human. Of the people preferring a chatbot, they emphasise the strength of the chatbot in that it does not \textit{judge} their responses when discussing personal topics~(P3, P14, P16, P27, P39, P52), for instance, an embarrassing medical question~(P27, P39, P44) and on relationships~(P2, P38). As exemplified by participant 3: \textit{``I would prefer to disclose more personal or sensitive topics to a chatbot rather than a human because it feels less judgmental and more private, allowing me to express myself freely without the fear of being evaluated''}. Of the six people preferring humans for disclosing, people refer to the lack of empathy within chatbots and the fear of their personal information becoming a part of the chatbot's data. As participant 20 explains regarding empathy: \textit{``No, there are not any topics I would disclose to a chatbot than to a human. I already know in advance the chatbot cannot feel empathy like a human so I would rather talk to a human, especially if it is a serious topic/discussion on something like mental health''}. For some, knowing that data typed into a chatbot may persist and not be subject to human forgetfulness may impact the information they are willing to disclose: \textit{``No, I don't want anything I say to get in its algorithm. Especially if it asks for personal or invasive information. A human might eventually forget what I say, but a chatbot will never forget''}~(P32).

\subsection{Exploratory Conversational Insights}
In addition to our findings, we present five exploratory insights derived from participants' conversations with the chatbot. To enhance our previous results, we include selected excerpts from these interactions. These excerpts highlight notable conversational dynamics between participants and the chatbot, offering concrete examples of the subtler positive aspects of using chatbots in surveys.

\subsubsection*{Participants Demanding More Specific Questions}
A key motivation for using chatbots in surveys is their potential to provide clarification when participants find questions unclear. Although this was not a dominant behaviour in our data, we did observe instances where participants directly engaged the chatbot to refine or specify questions they found too vague. In the example conversation (see Appendix~\ref{app_specific_question}), the participant perceives the initial question as too broad and requests further clarification through several steps. Eventually, the chatbot is able to narrow down the question to one the participant feels comfortable answering. This interaction highlights how the chatbot's conversational format can adapt to maintain relevance for both the study's objectives and the participant's needs.

\subsubsection*{Clarifying Technology Features as Part of the Evaluation Stage}
During the \tevaluation~stage of our study, we observed that participants, in addition to assessing the technology, would also ask the chatbot questions about its features. This helped them better understand how those features might impact their daily lives. In the example (see Appendix~\ref{app_clarifying_technology}), a participant inquired about the behaviour of calls when using the ``Do Not Disturb'' feature. After the chatbot's clarification, the participant was able to relate the explanation to their own experiences.

\subsubsection*{Returning to Past Points}
In several instances, participants mentioned multiple technology-related stressors in a single response. The chatbot was able to address these concerns individually, ultimately gaining deeper insights into those issues. In our example (see Appendix~\ref{app_past_points}), the participant expresses concerns about the importance of human presence and the potential overuse of AI in service roles. The chatbot first addresses the topic of human presence before transitioning to the AI concern, ensuring the participant has the opportunity to elaborate on both points.

\subsubsection*{Creating In-Conversation Realisations}
As a result of the conversational format of the survey, we observed participants making subtle realisations about their previous responses during the course of the interaction. In our example (see Appendix~\ref{app_creating_realisations}), one participant initially states that they use technology as an escape, but after several exchanges, returns to that statement, noting ``ironically'' how difficult it is to actually escape technology. This highlights a key advantage of conversational surveys over traditional open-ended questions, which may only capture immediate reactions. In contrast, the chatbot facilitates deeper reflection, enabling participants to refine or reconsider their responses as the conversation progresses.

\subsubsection*{Inquiring through Hypothetical Examples}
As researchers, we are particularly interested in our participants' personal opinions. However, participants may sometimes feel they have nothing relevant to contribute in certain areas of the study. In such cases, we observed that the chatbot was able to reframe questions, encouraging participants to consider how others might experience relief from technological stress. In the example (see Appendix~\ref{app_hypothetical_examples}), the chatbot and participant engage in a discussion of hypothetical scenarios aimed at gathering suggestions for improving current technologies, such as MS Office Excel. In contrast, a traditional open-ended question might have allowed the participant to move forward without offering any valuable insights.



\section{Discussion}
We investigated the effects of distinct and theory-based interview probes in chatbots for online HCI surveys on response quality and user perceptions. Using a custom chatbot, we explored four interview probes across three stages of HCI research. Our results show that the \tidiographic~probe performed the best overall in terms of response quality, while the \tidiographic, \tdescriptive, and \tclarifying~probes performed well across different stages, with the \tidiographic~consistently leading. Participants rated their chatbot experience similarly across probes, though the \texplanatory~probe was perceived as more repetitive. Qualitative findings revealed that the chatbot’s questioning style effectively encouraged reflection, potentially surpassing traditional open-ended questions. Participants preferred the conversational chatbot for qualitative insights and felt comfortable sharing information with both chatbots and humans, with chatbots favoured for personal thoughts due to their non-judgmental nature. Additionally, our exploratory insights highlighted notable conversational patterns facilitated by chatbots in online surveys. Next, we discuss the use of probes for qualitative data collection in HCI surveys, offering concrete recommendations for the use of specific probes at distinct research stages. We also address the challenges and opportunities for chatbots in online surveys and conclude with limitations and future directions.

\subsection{Interview Probes in HCI Surveys}
Interview probes are a well-established practice of qualitative research across various disciplines. Probes have even been used in a lightweight manner in surveys, where prior responses are embedded into subsequent items~\cite{oudejans2018using, Behr2012}. However, there has been limited exploration of how different questioning styles in chatbots influence data collection. Xiao et al.~\cite{Xiao2020a} demonstrated that \textit{active listening} techniques improved engagement and response quality. Building on this, we expand the range of strategies researchers may apply when using chatbots in surveys, drawing from the extensive literature on interview probes~\cite{Robinson2023}. Our approach aims to further enhance participant engagement~\cite{Heerwegh2007, Krosnick1999} and improve response quality in chatbot-driven surveys~\cite{Uka2021, Xiao2020, Ge2023, Kocielnik2021, Harms2017, Kim2019}. We found that participants were eager to engage with our chatbot and share their thoughts, opinions, and experiences. Consistent with previous research~\cite{Bickmore2011, Shamekhi2018}, we found that the conversational format not only enhanced engagement but also allowed participants to ask clarifying questions, narrow down broad topics, revisit previous points, and explore hypothetical scenarios. This interaction led participants to share more detailed and genuine~\cite{Li2017} information, including personal or sensitive insights, with the chatbot~\cite{Sundar2019, cox2022does}. Below we provide recommendations on which interview probes to use within the specific stages of HCI research, and summarised in Table~\ref{tab:meaure_probe_stage_summary}.

\begin{table}[b]
\caption{Overview of high-performing measures for each probe across the three interview stages.}
\label{tab:meaure_probe_stage_summary}
\begin{tabularx}{\linewidth}{@{}lXXX@{}}
\toprule
& \textbf{Exploration} & \textbf{Requirements} & \textbf{Evaluation} \\ \midrule
\textbf{Descriptive} & Relevance, Specificity & Relevance, Specificity, Clarity & --- \\
\textbf{Idiographic} & Relevance, Specificity & Informativeness, Relevance, Specificity, Clarity & Informativeness, Relevance, Specificity, Clarity \\
\textbf{Clarifying} & Specificity & Relevance, Clarity & --- \\
\textbf{Explanatory} & --- & --- & --- \\ 
\bottomrule
\end{tabularx}
\end{table}

\subsubsection*{Exploration Stage}
The exploratory stage focuses on understanding users' experiences, needs, and preferences with current or future technologies~\cite{lazar_research_2017}. Our results show that the \tdescriptive~and \tidiographic~probes were most effective for gathering rich, user-centred information at this stage. The \tdescriptive~probe encourages participants to share both their actions and thoughts. The probe performed well in relevance and specificity, suggesting that it is effective in capturing narratives that are both relevant to the interview’s goals and detailed enough to provide actionable insights. By helping users express both the emotional and situational aspects of their experiences, the \tdescriptive~probe is well-suited for gathering the rich narratives needed in this stage~\cite{gorden1987interviewing, Pasupathi2001}. The \tidiographic~probe focuses on eliciting specific, detailed memories. It showed strong performance in relevance, specificity and clarity, indicating that it may outperform other probes in eliciting useful information. By aiding participants in shifting from general to vivid recollections~\cite{Rottenberg2006, thomsen2009}, the \tidiographic~probe provides the depth needed to explore user behaviours and challenges, making it a valuable tool in this stage.

\subsubsection*{Requirements Stage}
In the requirements stage, where the goal is to gather detailed insights into users' needs and frustrations with current technologies~\cite{lazar_research_2017}, the \tidiographic~and \tdescriptive~probes were found to be the most effective in eliciting relevant and specific information. The \tidiographic~probe, grounded in techniques that encourage users to recall specific, vivid experiences~\cite{Brewer1986, Rottenberg2006, thomsen2009}, consistently performed well across informativeness, relevance, specificity, and clarity. This suggests that the \tidiographic~probe is particularly effective at prompting users to provide detailed examples of their needs and challenges, which are essential for defining actionable requirements. By guiding participants to move from general recollections to specific memories, the \tidiographic~probe is well-suited to capture the depth of information necessary for understanding the intricacies of users' experiences. The \tdescriptive~probe, which encourages participants to recount both the emotional and contextual aspects of their experiences~\cite{bruner1990acts, gorden1987interviewing, Pasupathi2001}, also demonstrated strong performance, particularly in relevance, specificity and clarity. This probe appears to be well-suited for gathering broad narratives that encompass both what users do and how they feel about their experiences. In the context of requirements gathering, the \tdescriptive~probe helps to capture not only the specific tasks or challenges users encounter but also the surrounding circumstances and motivations that inform their needs.

\subsubsection*{Evaluation Stage}
In the evaluation stage, where the primary goal is to gather feedback on prototypes or completed products to refine design and functionality~\cite{lazar_research_2017}, the \tidiographic~ probe provided the most valuable insights. The \tidiographic~probe demonstrated strong performance across informativeness, relevance, specificity, and clarity. By encouraging participants to recall specific, detailed memories~\cite{Brewer1986, Rottenberg2006}, the \tidiographic~probe is well-suited for uncovering nuanced feedback on user experiences with prototypes or products. This detailed recall may help highlight specific aspects of the product that need improvement or clarification, making the \tidiographic~probe a valuable tool during this stage. 

\subsection{Challenges and Opportunities for Chatbots in Surveys}
Our work builds on and extends prior studies that have shown chatbots as a viable solution to collect qualitative data~\cite{Uka2021, Xiao2020, Ge2023, Kocielnik2021, Harms2017, Kim2019}. Nevertheless, we also identified some challenges and new opportunities that warrant further investigation. 

\subsubsection*{Integrating Multiple Probes in Conversations}
Our research has demonstrated how different interview probes can yield varying quality results, as dependent on the specific research stage. Our findings are based on brief interactions, due to the survey format, where the chatbot and participant exchange several messages within a short time frame. A valuable avenue for future research is the integration of multiple interview probes within a single conversation, designed to span a longer time frame and engage more deeply with a topic~\cite{Robinson2023}. \textit{I.e.}, the \tdescriptive~probe could be employed to establish an initial topic for discussion, with the \tclarifying~probe asking for clarifications, and subsequently, the \texplanatory~probe supporting the participant in identifying relationships between their experiences and opinions.

\subsubsection*{Relinquishing Control in Surveys}
Deploying chatbots in surveys inherently results in reduced control over the study. LLM-based chatbots may behave unpredictably or deviate from the researcher’s intended line of questioning. Traditional surveys provide a structured, fixed path for large-scale data collection, but managing the dynamic nature of chatbot-driven interactions can pose challenges~\cite{heerwegh2008, roster2004, fricker2005}. 
However, this conversational format also presents an opportunity to gather richer qualitative data than enabled by traditional surveys. The use of theory-based interview probes, as explored in this paper, offers a way to maintain control while still fostering an engaging and flexible survey experience for participants~\cite{traum2017computational}.

\subsubsection*{Wrapping up Conversations or Topics}
The chatbot in our study did not enforce a specific endpoint for the conversation.
Instead, participants were instructed to type `goodbye' when they wished to move on. Despite the fact that participants controlled the start and end of the conversation, several participants felt the questions became repetitive, addressing points they had already covered. While clearer instructions could potentially mitigate this issue, we envision future opportunities to computationally assess conversations in real time, allowing the chatbot to suggest when it is appropriate to progress, rather than relying on participants to end the interaction. This highlights an important balance between achieving the desired conversational depth and avoiding participant fatigue~\cite{porter2004overcoming, Lavrakas2008, Porter2004a, Porter2004b}. 

This is not without challenges, as assessing whether a participant has provided the extent of detail they are able and willing to share, or whether further probing might uncover additional insights is inherently difficult---including in human-driven interviews. To address this, we propose three distinct opportunities for researchers to explore. First, one potential direction is to determine the conclusion of a chatbot conversation by evaluating whether participant responses meet predefined quality criteria. For instance, an additional agent could evaluate each response to determine whether it sufficiently addresses the chatbot's questions. Second, a more complex solution involves conducting a computational analysis of the entire conversation. This approach aligns with the broader challenge of addressing data saturation~\cite{Morse1995, Fusch2015} (or data adequacy) within qualitative research. While computational methods may offer the capability to predict a level of saturation and guide the conversation to a conclusion, this foundational challenge may be out of scope to resolve in an automated fashion. Third, we propose that an agent could evaluate a decreasing trend in the quality of participant responses. For example, this could involve analysing responses for signs of repetitiveness, reduced answer length, or a diminishing level of detail, which might indicate a decline in engagement or cognitive fatigue~\cite{karim2024}. These suggestions are, of course, non-exhaustive, but offer potential trajectories for dynamically determining the termination of chatbot conversations in surveys.

\subsubsection*{Contextuality is Conserved in the Conversation}
A key observation from our study is the nuanced richness provided by the chat data. Similar to face-to-face interviews, the most valuable insights often lie not in individual participant statements but in the broader context of the conversation. We regularly observed that participants struggled to articulate their thoughts early on in the chat, but that the step-by-step probing gradually elicited deeper insights, sometimes even to the surprise of the participants themselves. These reflections, which may have been lost without probing, underscore the potential for chatbots in the context of surveys. However, for effective conversational data, the participant's initial responses must provide sufficient detail to enable the chatbot to generate meaningful follow-up probing questions~\cite{mctear2023, dam2024}. Without this foundation, the quality of the probing and the depth of insights may be limited, highlighting the importance of designing chatbot interactions that encourage detailed initial participant responses.

\subsection{Limitations and Future Work}
We acknowledge several limitations in our study. In this study, we investigated the utilisation of theory-based interview probes for qualitative data collection in surveys through chatbots, however, comparing the effectiveness of surveys with LLM-based chatbots with traditional open-ended questions fell outside of our scope. Future research could investigate the quality of responses generated by LLM-based chatbots as compared to traditional open-ended survey questions to better understand their relative strengths and limitations, further extending the work of~\citet{Xiao2020}. This line of investigation could include both qualitative and quantitative approaches. For instance, manual thematic analysis could be employed to uncover patterns in the data, such as the depth and richness of the responses, the coherence of narratives, or the relevance of ideas to the questions posed. Computational analyses, on the other hand, could provide complementary insights by systematically evaluating various quality dimensions, such as response diversity, originality, sentiment, and informativeness, as exemplified by metrics like the diversity of ideas~\cite{cox2021}. Furthermore, comparisons could be expanded to evaluate how specific enhancements, such as interview probes, influence the quality of data when used in conjunction with state-of-the-art LLMs. For example, these comparisons could examine whether probes elicit more detailed, contextually relevant, or actionable responses compared to pre-defined chatbot interactions. Studies might also investigate how participants perceive the clarity and supportiveness of probe-augmented chatbots versus traditional approaches or simpler LLM-driven systems without additional instructions. By addressing these questions, future research could not only assess the added value of interview probes but also offer practical recommendations for designing chatbots tailored to specific research goals. This would help advance the use of LLM-based tools in domains requiring high-quality qualitative data collection.

Second, we chose to focus on the phenomenon of technologically induced stress as the survey's topic, which provided a useful context for examining the interview probes. However, we recognise that this topic may not be representative of broader areas in HCI research. For instance, in the exploratory stage, researchers may not yet have a well-defined research problem, making depth-seeking interview probes less applicable. Additionally, while chatbots may be advantageous for evaluating specific design features in online surveys, the use of interview probes in this context may not generate the intended insights for guiding future design directions. Future work could, therefore, investigate the use of interview probes in different areas of HCI.

Third, we observed that the evaluation stage both elicited fewer responses and resulted in lower scores on the quality measures. We believe this may be due to two factors. One possibility is participant fatigue, as the evaluation stage occurs at the end of the process, potentially leading to diminished engagement. Alternatively, the lower response quality may reflect the inherent limitations in the insights participants can offer during the evaluation stage. Regardless, it is essential for future research to further investigate the dynamics of the evaluation stage, particularly in the context of using chatbots in surveys.



\section{Conclusion}
We developed an LLM-based chatbot to investigate the use of theory-based interview probes for collecting qualitative data in online surveys across three stages of HCI research. 
Specifically, we evaluated \tdescriptive, \tidiographic, \tclarifying, and \texploration~probes across the stages of \texploration, \trequirements, and \tevaluation.
Our results show that the \tidiographic~and \tdescriptive~probes provide high-quality insights in the \texploration~and \trequirements~stage, and that the \tidiographic~ probe performs well in the \tevaluation~stage. 
We found no larger experiential differences between the different probes, but uncovered several perceptions on the use of chatbots in service including exploratory conversational patterns. We provide concrete recommendations for the use of chatbots with interview probes in online surveys and critically discuss further challenges and opportunities on how to continue research into this promising way for online survey data collection. 


\begin{acks}
This work is supported by the Carlsberg Foundation, grant CF21-
0159.
\end{acks}

\bibliographystyle{ACM-Reference-Format}





\appendix
\section{Appendix}
\subsection{Tasks}\label{appen_tasks}
Below we provide the informational text for each task that was shown to the participants for each of the three stages. In Table~\ref{tab:tasks}, each image/video for each research stage is shown along side the description.

\subsubsection{Exploration} For this first stage, we are interested in understanding your overall experiences with technology and how it affects your well-being. You will be shown three different images depicting situations where technology may be part of your daily life. Afterward, you will be asked questions about these situations.

\subsubsection{Requirements} For this stage, the goal is to understand the desires and expectations you have for improving current or future technologies. We want to learn about your needs, goals, and frustrations with the technologies you use in your life. You will be shown three different images depicting situations where technology may be part of your day. Afterward, you will be asked questions about these situations.

\subsubsection{Evaluation} In this stage, we will collect your feedback on a specific solution for managing technology use: the 'Do Not Disturb' feature for Apple iOS. You will now be shown a video about this feature. Please watch the video in its entirety. While watching, please consider how this feature may impact your life and the technologies you use.

\begin{table*}[h]
\begin{tabularx}{\linewidth}{@{}>{\raggedright\arraybackslash}m{0.14\linewidth}
                                >{\raggedright\arraybackslash}m{0.26\linewidth}
                                >{\raggedright\arraybackslash}m{0.26\linewidth}
                                >{\raggedright\arraybackslash}m{0.26\linewidth}@{}}
\toprule
    & \textbf{Home} & \textbf{Public} & \textbf{Work} \\ 
    \midrule
    \textbf{Exploration} & & & \\
    \textit{Image shown} &
    \includegraphics[width=\linewidth]{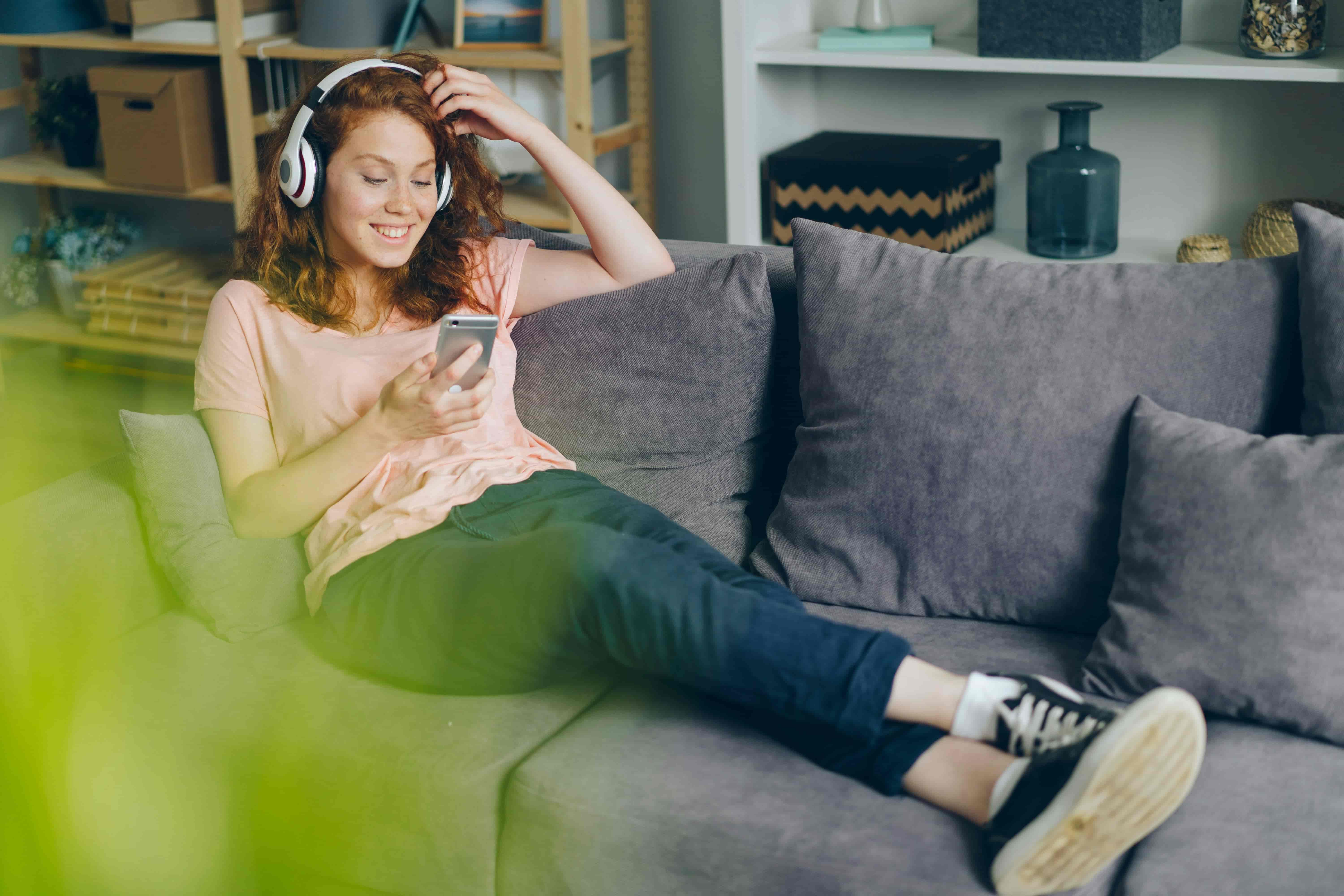} &
    \includegraphics[width=\linewidth]{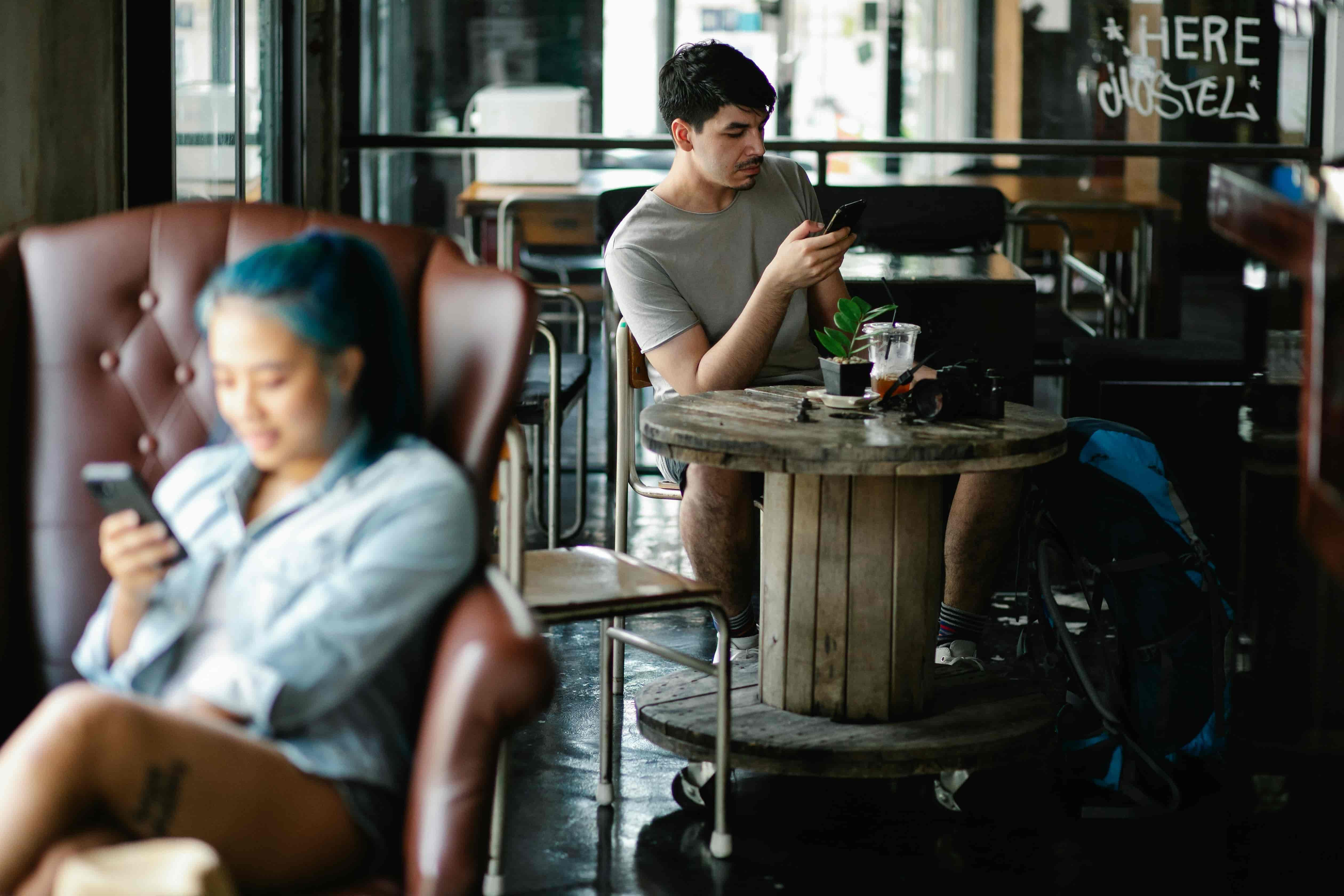} & 
    \includegraphics[width=\linewidth]{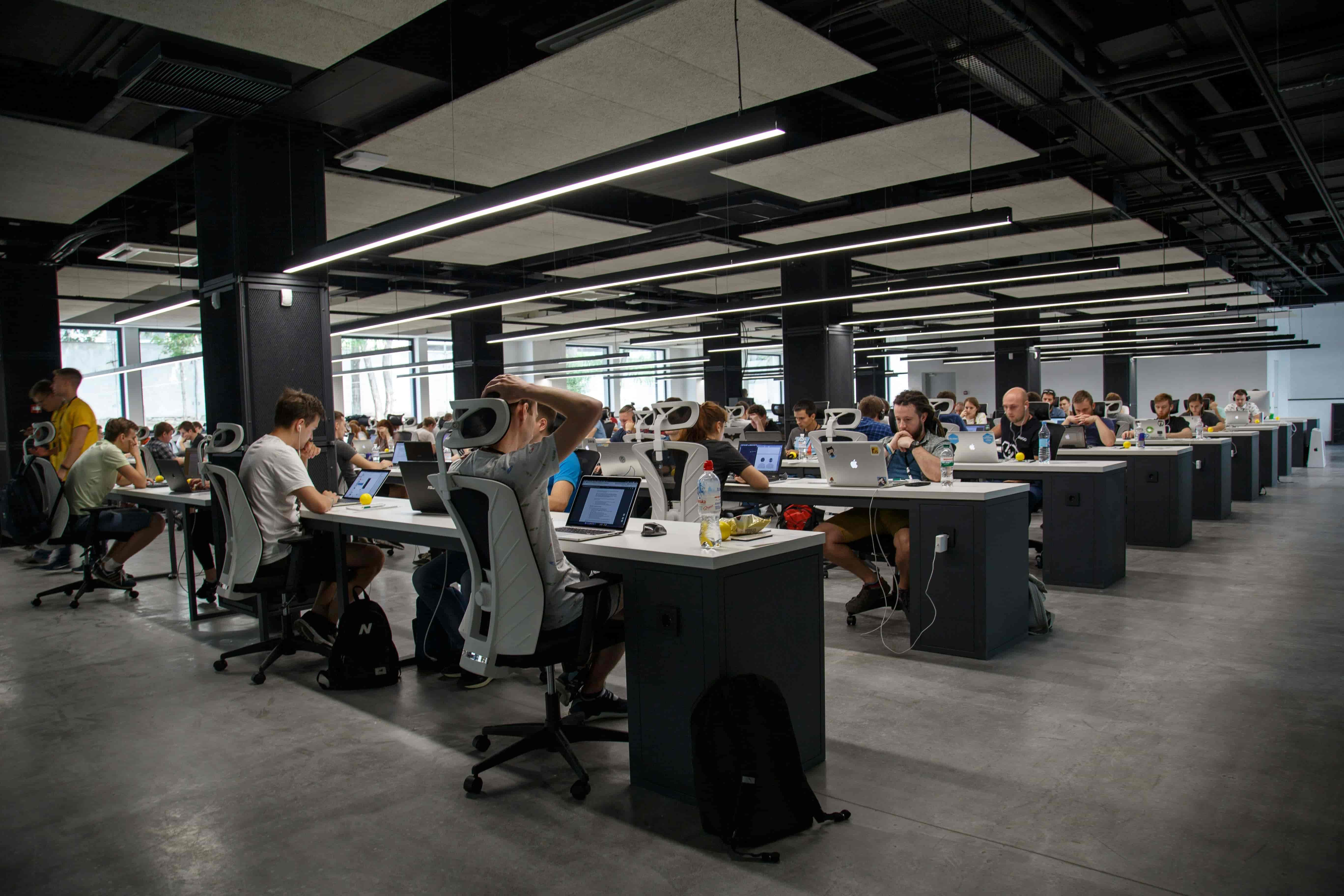} \\  
    \textit{Text description} & 
    Consider how different technologies affect your life when you are at home, and how this impacts both you and the people around you. & Consider how technology affects your experiences and activities when you are outside in public. & Consider how your work life is affected by technology, including situations where you use it or are influenced by it. \\
    \textbf{Requirements} & & & \\
    \textit{Image shown} &
    \includegraphics[width=\linewidth]{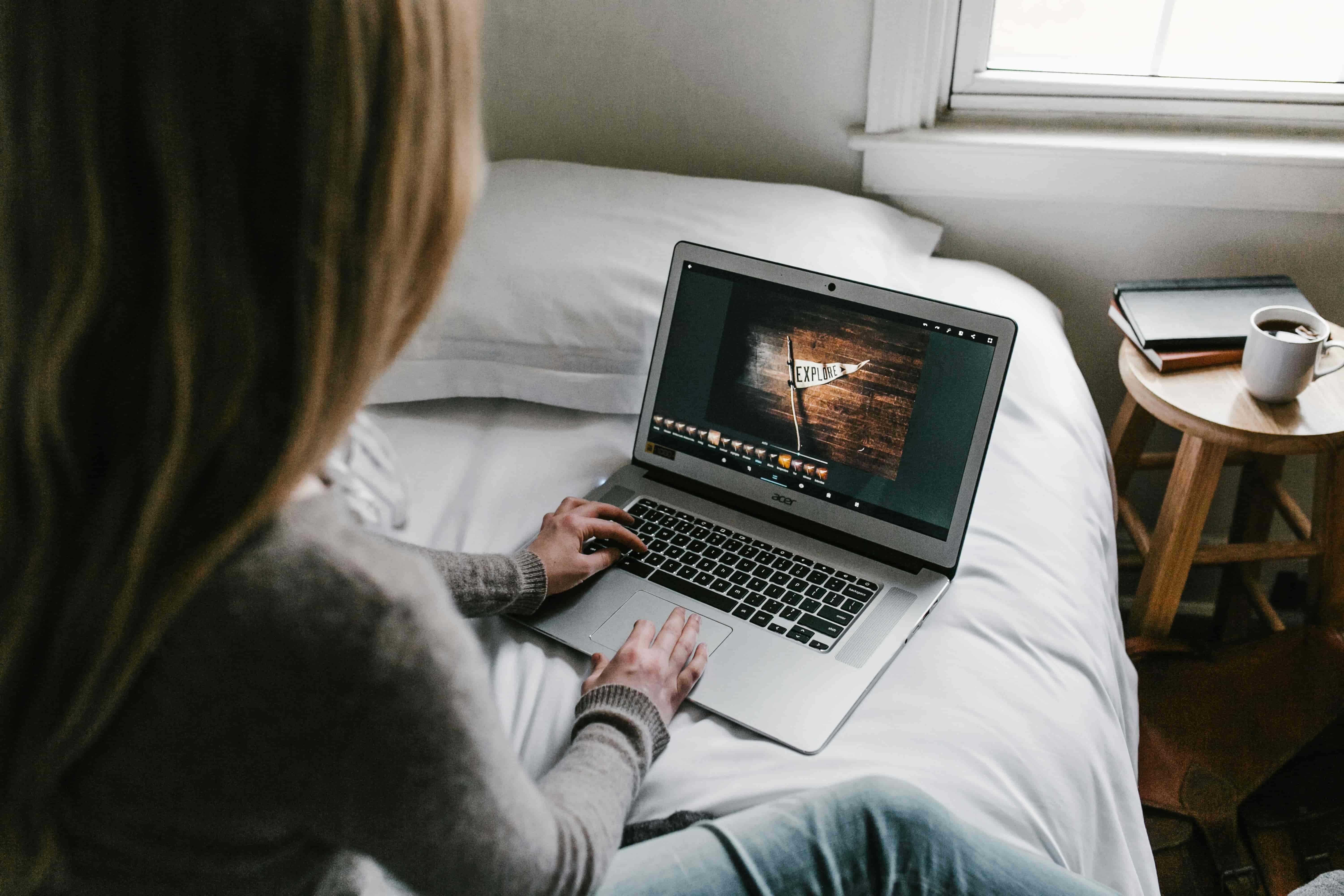} &
    \includegraphics[width=\linewidth]{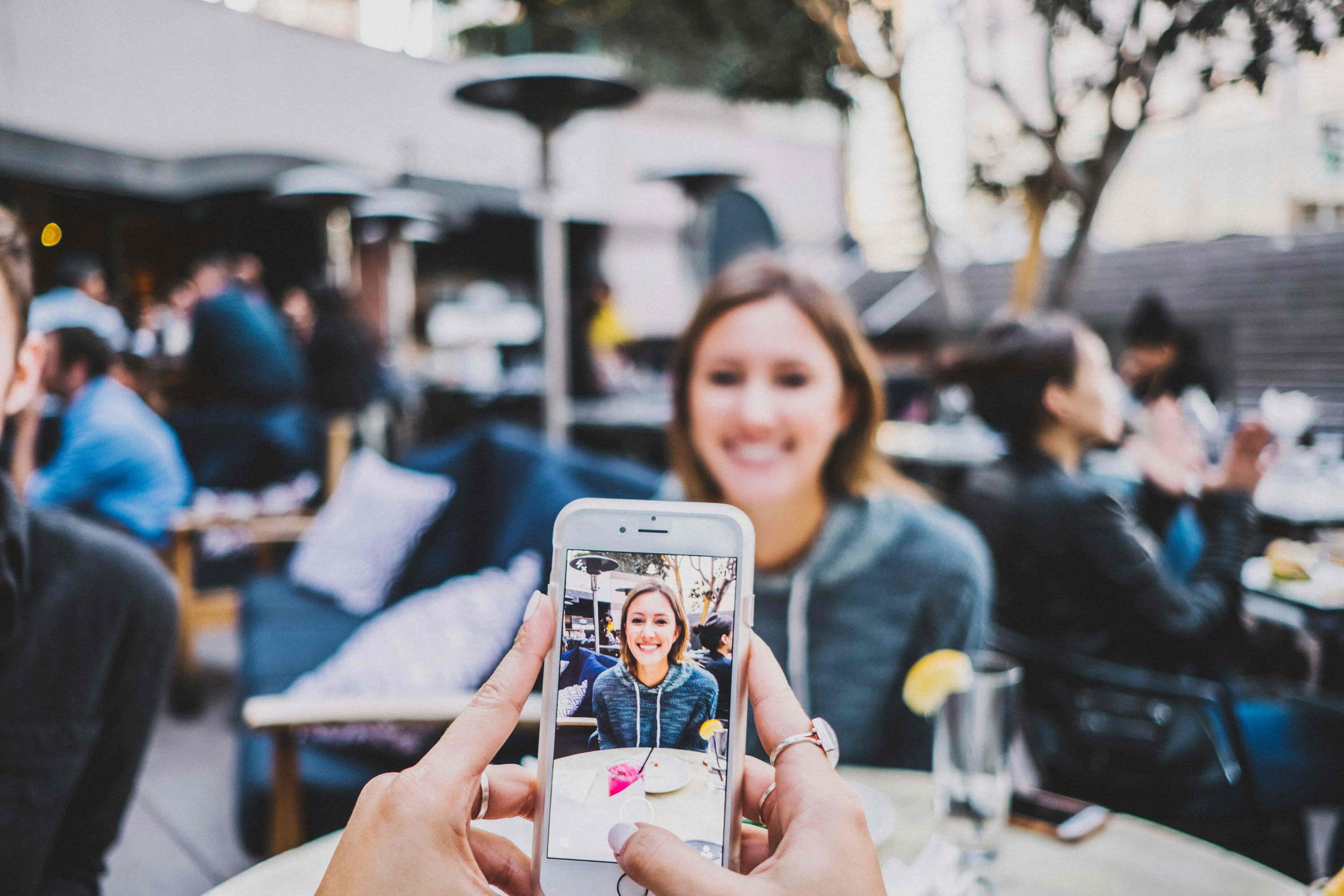} & 
    \includegraphics[width=\linewidth]{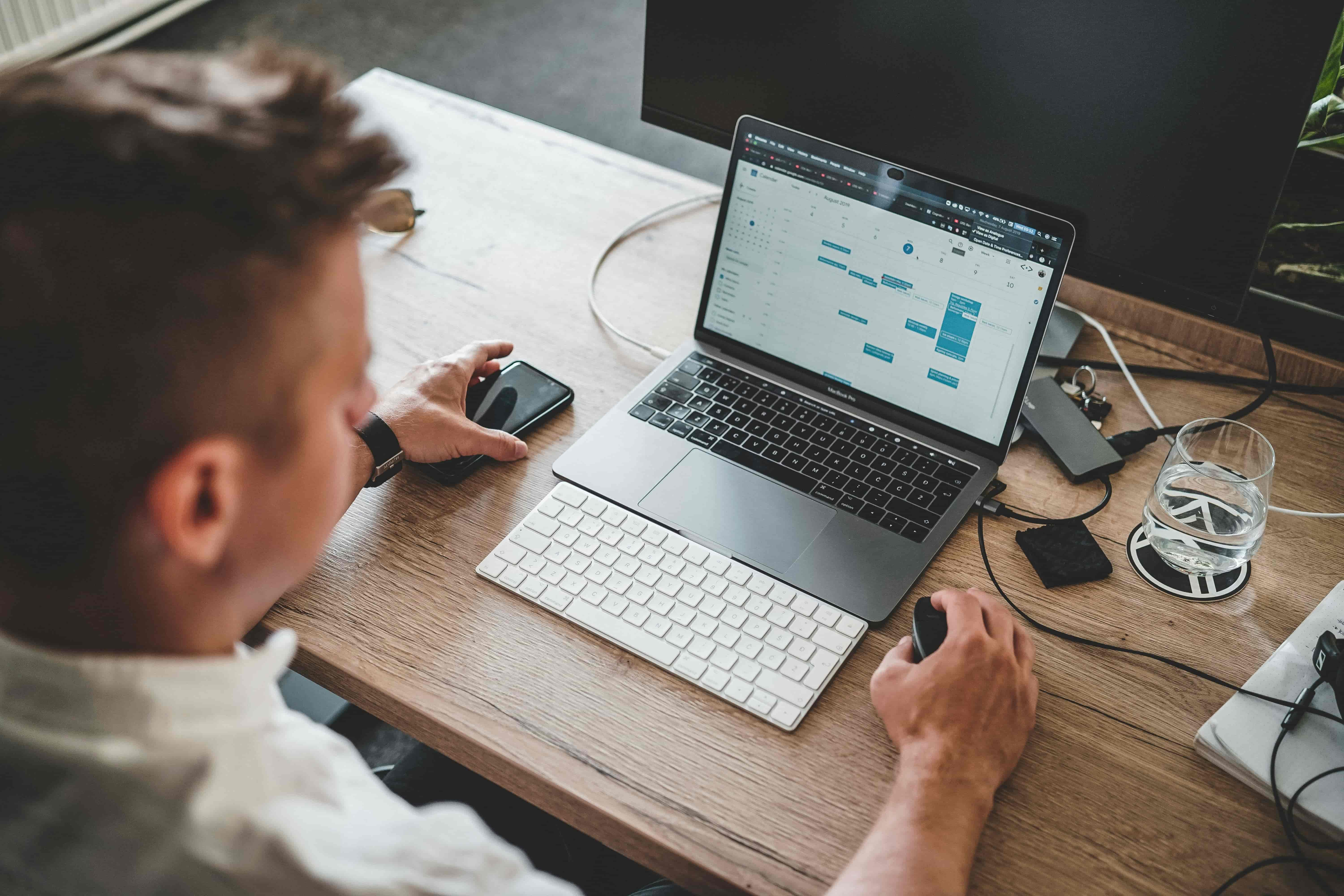} \\  
    \textit{Text description} & 
    Consider how your life when at home is affected by technology, and think about what improvements could be made to current or future technology in a positive way. & Consider when you are outside in public, and think about what improvements could be made to current or future technology in a positive way. & Consider how your work life is being affected by technology, and what can be done to current or future technology in a positive way. \\
    \textbf{Evaluation} & & & \\
    \textit{Video} & & \includegraphics[width=\linewidth]{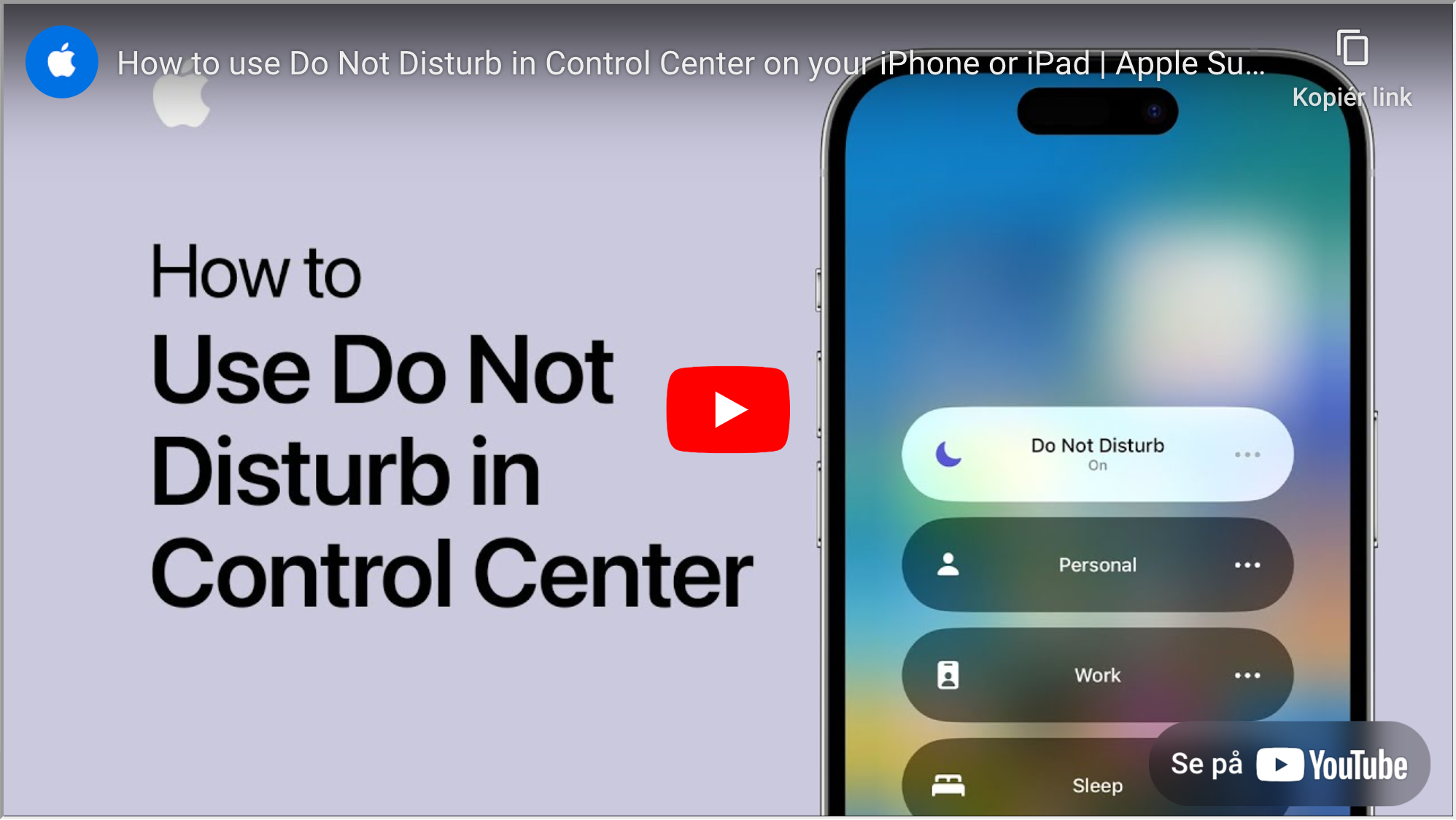} & \\
    
\end{tabularx} 
\caption{Table showing the different images/video shown for each research stage.}
\label{tab:tasks}
\end{table*}

\subsection{Example Responses for the Gricean Maxims}\label{example_responses}
\subsubsection{Relevance}
\underline{Relevance score of 0}:
``It is a tool, and should be treated as such''

\underline{Relevance score of 1}:
``just random posts or news''

\underline{Relevance score of 2}:
``It changed my mood for the rest of the afternoon because my mind went to work when it should have been on my family. I was upset at myself for checking my phone and taking my time away.''

\subsubsection{Specificity}

\underline{Specificity score of 0}:
``Yes, I have''

\underline{Specificity score of 1}:
``I don't want to get sidetracked from what is really important.''

\underline{Specificity score of 2}:
``Recently, I was away for the weekend with my family enjoying quiet time at the lake. I checked my phone and saw I had a work related email that needed my attention. I spent 30 minutes away from my family while I went to handle the situation. I was upset and stressed that I had to do this on a Saturday while I should have been enjoying fun time with my family.''

\subsubsection{Clarity}
\underline{Clarity score of 0}:
``inconvebient'' and ``NA''

\underline{Clarity score of 1}:
``Sticking with what works''

\underline{Clarity score of 2}:
``At work, they have many different platforms that they use for communication, meetings, tracking training, tracking schedules, calendars, etc. They also keep updating the program we use for patient care charting so there are new or different boxes to fill out all of the time.''

\subsection{Calculating Smoothness from SEQ}\label{seq_calc}
\[
    smoothness = \frac{easy + (8-tense) + pleasant + smooth + (8-uncomfortable)}{5} 
\]

\subsection{Chat Excerpts}
\subsubsection{Participants Demanding More Specific Questions}\label{app_specific_question}

\aptLtoX{\begin{description}
\item \Chatbot Hello, I would like to discuss the impact of technology on stress. Have you experienced technology having a stressing effect on your life?
\item \User Sometimes
\item \Chatbot Thank you for sharing that. Can you explain when you typically experience this stress and what usually triggers it?
\item \User First, give me a few examples
\item \Chatbot Of course. For example, some people feel stressed due to a high volume of work emails, constant notifications from apps, difficulties in learning new software, or pressure to stay connected online all the time. Do any of these resonate with your experiences?
\item \User No
\item \Chatbot Thank you for clarifying. Can you explain what specific situations or interactions with technology cause you stress?
\item \User It's still hard to answer that since it is a broad question that could mean anything
\item \Chatbot I understand. Let's narrow it down. Can you describe a recent situation where interacting with technology left you feeling stressed or overwhelmed? What specifically caused that feeling in that instance?
\item \User Narrow it down further please
\item \Chatbot Absolutely. Have you ever felt stressed when using your smartphone specifically? If so, what in particular caused that stressâ€”such as a specific app, feature, or notification?
\item \User Over my phone, yes. So much that it would sound overrated calling it a 'smartphone' since it's not as smart as industries claim it to be. It's more like an app phone or feature phone and nothing else. Considering how every app and photos / videos takes up huge chunks of storage space and the battery life is terrible and faulty. For example, I would have to constantly clear up space for any new apps I'd want to install. Another example is how my phone would instantly turn itself off when my battery reaches 30\% battery while playing a game. No low battery indicator warnings or anything. And the app store is mostly junk apps that are filled with crappy games and poorly made applications infested with ads and in-app purchases just to get it to work.
\end{description}}{\begin{description}
\Chatbot Hello, I would like to discuss the impact of technology on stress. Have you experienced technology having a stressing effect on your life?
\User Sometimes
\Chatbot Thank you for sharing that. Can you explain when you typically experience this stress and what usually triggers it?
\User First, give me a few examples
\Chatbot Of course. For example, some people feel stressed due to a high volume of work emails, constant notifications from apps, difficulties in learning new software, or pressure to stay connected online all the time. Do any of these resonate with your experiences?
\User No
\Chatbot Thank you for clarifying. Can you explain what specific situations or interactions with technology cause you stress?
\User It's still hard to answer that since it is a broad question that could mean anything
\Chatbot I understand. Let's narrow it down. Can you describe a recent situation where interacting with technology left you feeling stressed or overwhelmed? What specifically caused that feeling in that instance?
\User Narrow it down further please
\Chatbot Absolutely. Have you ever felt stressed when using your smartphone specifically? If so, what in particular caused that stressâ€”such as a specific app, feature, or notification?
\User Over my phone, yes. So much that it would sound overrated calling it a 'smartphone' since it's not as smart as industries claim it to be. It's more like an app phone or feature phone and nothing else. Considering how every app and photos / videos takes up huge chunks of storage space and the battery life is terrible and faulty. For example, I would have to constantly clear up space for any new apps I'd want to install. Another example is how my phone would instantly turn itself off when my battery reaches 30\% battery while playing a game. No low battery indicator warnings or anything. And the app store is mostly junk apps that are filled with crappy games and poorly made applications infested with ads and in-app purchases just to get it to work.
\end{description}}

\subsubsection{Clarifying Technology Features as Part of the Evaluation Phase} \label{app_clarifying_technology}

\aptLtoX{
\begin{description}
\item \Chatbot Got it. So, prioritizing checking emails on your own schedule rather than being interrupted by alerts might be beneficial. Could you describe any potential downsides or challenges you might foresee with leaving "do not disturb" on all the time?
\item \User I might not see calls and texts? Do calls still come through with it on?
\item \Chatbot Calls are silenced when "Do Not Disturb" is on. Given that, how would missing potentially important calls or texts impact your stress levels or daily routine?
\item \User I would want to silence notifications but not calls. There's probably a way to do that.
\item \Chatbot Yes, you can customize "Do Not Disturb" settings to allow calls from certain people. Could you clarify how having this level of control over notifications might influence your stress levels with technology?
\item \User Less junk taking my attention away from what I am concentrating on would help stress levels for me.
\end{description}}{
\begin{description}
\Chatbot Got it. So, prioritizing checking emails on your own schedule rather than being interrupted by alerts might be beneficial. Could you describe any potential downsides or challenges you might foresee with leaving "do not disturb" on all the time?
\User I might not see calls and texts? Do calls still come through with it on?
\Chatbot Calls are silenced when "Do Not Disturb" is on. Given that, how would missing potentially important calls or texts impact your stress levels or daily routine?
\User I would want to silence notifications but not calls. There's probably a way to do that.
\Chatbot Yes, you can customize "Do Not Disturb" settings to allow calls from certain people. Could you clarify how having this level of control over notifications might influence your stress levels with technology?
\User Less junk taking my attention away from what I am concentrating on would help stress levels for me.
\end{description}}

\subsubsection{Returning to Past Points} \label{app_past_points}
After the participant was being asked about some possible ways to change technology to better fit your needs, and how they could believe technology could be improved to decrease its stressing effects on their life. 

\aptLtoX{\begin{description}
\item \User Apps that remind users to stay in the present and interact with people through other means than just technology. Stop giving customer service jobs to AI bots. Always allow pressing zero to talk to a human.
\item \Chatbot Thank you for sharing that. Regarding apps that remind users to stay present, can you clarify what kind of reminders or features you would find most helpful in such an app?
\item \User Maybe like daily affirmations sent via SMS or within an app to remind people to partake in human interaction. Be kind, compliment someone, strike up a conversation with someone you might not otherwise talk to.
\item \Chatbot I see, daily affirmations encouraging human interaction sound beneficial. Can you provide an example of a situation where receiving such a reminder would have significantly impacted your day positively?
\item \User Not specifically, I think it would just be nice for people to have that reminder once in a while so they don't get so focused on their screens, particularly when out in public.
\item \Chatbot Got it. Shifting to your point about customer service AI bots, can you explain in more detail why you find AI bots stressful and how this stress arises?
\item \User AI is taking over too many jobs that people need. If you take away lower level jobs from people who don't have any education or experience beyond what they're already doing for work, it becomes a problem when that person can't afford new training and/or provide for their family.
\end{description}}{
\begin{description}
\User Apps that remind users to stay in the present and interact with people through other means than just technology. Stop giving customer service jobs to AI bots. Always allow pressing zero to talk to a human.
\Chatbot Thank you for sharing that. Regarding apps that remind users to stay present, can you clarify what kind of reminders or features you would find most helpful in such an app?
\User Maybe like daily affirmations sent via SMS or within an app to remind people to partake in human interaction. Be kind, compliment someone, strike up a conversation with someone you might not otherwise talk to.
\Chatbot I see, daily affirmations encouraging human interaction sound beneficial. Can you provide an example of a situation where receiving such a reminder would have significantly impacted your day positively?
\User Not specifically, I think it would just be nice for people to have that reminder once in a while so they don't get so focused on their screens, particularly when out in public.
\Chatbot Got it. Shifting to your point about customer service AI bots, can you explain in more detail why you find AI bots stressful and how this stress arises?
\User AI is taking over too many jobs that people need. If you take away lower level jobs from people who don't have any education or experience beyond what they're already doing for work, it becomes a problem when that person can't afford new training and/or provide for their family.
\end{description}}

\subsubsection{Creating In-Conversation Realisations} \label{app_creating_realisations}

\aptLtoX{\begin{description}
\item \User Sometimes, I use technology to escape reality. But in all honesty, sometimes not having that escape causes me anxiety. I probably wouldn't struggle with that anxiety if I had a healthier coping mechanism to deal with day-to-day stress that makes me feel the need to escape.
\item \Chatbot I appreciate your openness. Could you elaborate on why you feel the absence of that technological escape increases your anxiety?
\item \Removed [16 messages back and fourth between the participant and the chatbot in between here]
\item \User I think it has helped a little. But I still think that technology overall has such a big influence on our lives, and is almost impossible to do away with entirely that there's really no escaping it... Ironically, I mentioned first that I use technology as an escape, but I suppose the reality is that there is no escaping technology.
\end{description}}{
\begin{description}
\User Sometimes, I use technology to escape reality. But in all honesty, sometimes not having that escape causes me anxiety. I probably wouldn't struggle with that anxiety if I had a healthier coping mechanism to deal with day-to-day stress that makes me feel the need to escape.
\Chatbot I appreciate your openness. Could you elaborate on why you feel the absence of that technological escape increases your anxiety?
\Removed [16 messages back and fourth between the participant and the chatbot in between here]
\User I think it has helped a little. But I still think that technology overall has such a big influence on our lives, and is almost impossible to do away with entirely that there's really no escaping it... Ironically, I mentioned first that I use technology as an escape, but I suppose the reality is that there is no escaping technology.
\end{description}}

\subsubsection{Inquiring through Hypothetical Examples} \label{app_hypothetical_examples}
After the participant was being asked about some possible ways to change technology to better fit your needs, and how they could believe technology could be improved to decrease its stressing effects on their life. 

\aptLtoX{\begin{description}
\item \User I don't see how it can be improved relating to the stress it might cause people. I don't get stressed by technology.
\item \Chatbot Thanks for sharing that. Even if you donâ€™t personally experience stress from technology, can you think of situations where others might feel stressed by it? Could you describe a specific scenario where someone might feel overwhelmed by technology?
\item \User A person unfamiliar with it at a new job would be overwhelmed, I'm sure. They would require the correct training by a patient instructor. If an app on the device were installed that taught them how to use the device for the task they were working on, that would be helpful.
\item \Chatbot Thank you for that scenario. Can you describe in more detail what aspects of using new technology might overwhelm someone in a new job? For instance, what specific tasks or features might cause confusion or stress?
\item \User If someone had never used Excel, they would have no idea where to begin. They definitely would need prior instruction.
\item \Chatbot That makes sense. Can you give an example of how a lack of knowledge in Excel specifically might impact someone's workday? What kinds of problems or stress might they encounter because of this?
\item \User If they lied when they were hired for the job and said they were proficient in Excel, then were given an assignment on their first day to produce a spreadsheet detailing the company's finances, that would be stressful. I think it would ruin their work day. 
\item \Chatbot Thanks for that detailed example. In such a stressful situation, what kind of support or resources do you think would make the most significant difference for that person? Can you describe any particular tools or assistance that could help alleviate their stress?
\item \User If there were instruction in the Excel app, when you clicked one of the functions, if it gave more information on how to use it, that might be helpful. Or an instructional video inside the app would help.
\item \Chatbot That sounds like a useful solution. Can you elaborate on how you envision the instructional content being integrated into the Excel app? For example, what specific features or types of information would be most helpful in reducing stress for the user?
\item \User If there were a separate part of Excel, if when you opened the application, you have a choice to select "instruction" or a manual.
\item \Chatbot I see. Could you describe in more detail how this instructional section might look and operate? For instance, what kind of content should it include and how would it guide users through learning Excel?
\item \User It could be divided into the various functions of Excel. A quick short video showing what each button is for, and the capabilities of it. Like a short "chapter" for each function, either text or video.
\end{description}}{
\begin{description}
\User I don't see how it can be improved relating to the stress it might cause people. I don't get stressed by technology.
\Chatbot Thanks for sharing that. Even if you donâ€™t personally experience stress from technology, can you think of situations where others might feel stressed by it? Could you describe a specific scenario where someone might feel overwhelmed by technology?
\User A person unfamiliar with it at a new job would be overwhelmed, I'm sure. They would require the correct training by a patient instructor. If an app on the device were installed that taught them how to use the device for the task they were working on, that would be helpful.
\Chatbot Thank you for that scenario. Can you describe in more detail what aspects of using new technology might overwhelm someone in a new job? For instance, what specific tasks or features might cause confusion or stress?
\User If someone had never used Excel, they would have no idea where to begin. They definitely would need prior instruction.
\Chatbot That makes sense. Can you give an example of how a lack of knowledge in Excel specifically might impact someone's workday? What kinds of problems or stress might they encounter because of this?
\User If they lied when they were hired for the job and said they were proficient in Excel, then were given an assignment on their first day to produce a spreadsheet detailing the company's finances, that would be stressful. I think it would ruin their work day. 
\Chatbot Thanks for that detailed example. In such a stressful situation, what kind of support or resources do you think would make the most significant difference for that person? Can you describe any particular tools or assistance that could help alleviate their stress?
\User If there were instruction in the Excel app, when you clicked one of the functions, if it gave more information on how to use it, that might be helpful. Or an instructional video inside the app would help.
\Chatbot That sounds like a useful solution. Can you elaborate on how you envision the instructional content being integrated into the Excel app? For example, what specific features or types of information would be most helpful in reducing stress for the user?
\User If there were a separate part of Excel, if when you opened the application, you have a choice to select "instruction" or a manual.
\Chatbot I see. Could you describe in more detail how this instructional section might look and operate? For instance, what kind of content should it include and how would it guide users through learning Excel?
\User It could be divided into the various functions of Excel. A quick short video showing what each button is for, and the capabilities of it. Like a short "chapter" for each function, either text or video.
\end{description}}

\end{document}